%% file: hfs-paper-4.tex
\documentclass[fleqn,usenatbib]{mnras}

\usepackage{newtxtext,newtxmath}

\usepackage[T1]{fontenc}

\DeclareRobustCommand{\VAN}[3]{#2}
\let\VANthebibliography\thebibliography
\def\thebibliography{\DeclareRobustCommand{\VAN}[3]{##3}\VANthebibliography}

\usepackage{graphicx}	
\usepackage{amsmath}	
\usepackage{amssymb}	
\usepackage{hyperref}
\usepackage{float}
\usepackage{subfig}
\usepackage[detect-weight=true,per-mode=power]{siunitx}
\usepackage{xcolor}

\newcommand{\figref}[2][Figure~]{#1\ref{#2}}
\newcommand{\tabref}[2][Table~]{#1\ref{#2}}
\newcommand{\secref}[2][Section~]{#1\ref{#2}}
\renewcommand{\eqref}[2][equation~]{#1\ref{#2}}
\newcommand{\appref}[2][Appendix~]{#1\ref{#2}}

\DeclareSIUnit\Msol{M_{\odot}} 
\DeclareSIUnit\Lsol{L_{\odot}} 
\DeclareSIUnit\parsec{pc} 
\DeclareSIUnit\Jy{Jy} 
\DeclareSIUnit\beam{beam} 
\DeclareSIUnit\micron{\micro\metre} 
\DeclareSIUnit\pixel{px} 
\DeclareSIUnit\gauss{G} 
\DeclareSIUnit\sc{\centi\metre\squared} 
\DeclareSIUnit\cc{\centi\metre\cubed} 
\DeclareSIUnit\yr{yr} 

\newcommand{\nthp}{N\textsubscript{2}H\textsuperscript{+}} 
\newcommand{\molh}{H\textsubscript{2}} 
\newcommand{\vlsr}{$V_\mathrm{LSR}$} 
\newcommand{\mwydyn}{\textsc{mwydyn}} 

\usepackage{tipa} 

\usepackage{xcolor} 

\defcitealias{Anderson2021}{Paper I}

\title[ALMA study of hub-filament systems]{An ALMA study of hub-filament systems\\II. Quiescent filaments converging towards highly dynamic hubs}

\author[M. Anderson et al.]{
Michael Anderson,$^{1}$
Nicolas Peretto,$^{1}$\thanks{E-mail: PerettoN@cardiff.ac.uk}
Sarah E. Ragan,$^{1}$
Andrew J. Rigby,$^{2}$
Adam Avison,$^{3,4,5}$
\newauthor{}
Ana Duarte-Cabral,$^{1}$
Gary A. Fuller,$^{3,4,6}$
Yancy L. Shirley,$^{7}$
Alessio Traficante$^{8}$
and Gwenllian M. Williams$^{2,9}$
\\
$^{1}$ Cardiff Hub for Astrophysics Research and Technology (CHART), School of Physical, Chemical and Environmental Sciences, Cardiff University, Cardiff CF24 3AA, UK\\
$^{2}$School of Physics and Astronomy, University of Leeds, Leeds LS2 9JT, UK\\
$^{3}$ Jodrell Bank Centre for Astrophysics, Department of Physics and Astronomy, School Of Natural Science, The University of Manchester, Manchester M13 9PL, UK\\
$^{4}$UK ALMA Regional Centre Node, Manchester M13 9PL, UK\\
$^{5}$SKA Observatory, Jodrell Bank, Lower Withington, Macclesfield SK11 9FT, UK\\
$^{6}$I. Physikalisches Institut, University of Cologne, Z{\"u}lpicher Str 77, D-50937 K{\"o}ln, Germany\\
$^{7}$Steward Observatory, University of Arizona, 933 North Cherry Avenue, Tucson, AZ, 85721 USA\\
$^{8}$IAPS-INAF, Via Fosso del Cavaliere, 100, I-00133, Rome, Italy\\
$^{9}$Department of Physics, Aberystwyth University, Ceredigion SY23 3BZ, UK
}

\date{Accepted 26/10/2026. Received 26/10/2026; in original form 03/07/2026}

\pubyear{2024}

\begin{document}
\label{firstpage}
\pagerange{\pageref{firstpage}--\pageref{lastpage}}
\maketitle

\begin{abstract}
	Hub-filament systems are networks of converging interstellar filaments, often with active star formation at their centres, that may play an important role in high-mass star formation. In the first paper of this series, we found that the mass fraction that ends up in a clump’s most massive core is significantly higher in IR-dark hubs than IR-bright clumps, suggesting that the most-massive cores form early on. Such early massive core formation requires large inflow rates and dynamically active IR-dark clumps. We now present \nthp(J=1--0) observations of six IR-dark hub-filament systems mapped with ALMA 12m, 7m, and total power at $\sim3\si{\arcsecond}$ resolution, to trace the kinematics of the dense gas. The data show intricate emission structures and complex spectra. To characterise their kinematics, we have developed \mwydyn{}, a fully-automated, multiple velocity component, hyperfine line-fitting code. Our results reveal that the emission invariably consists of quiescent individual filaments in the outskirts that converge towards the hub centres where a systematic increase in velocity dispersion and number of components is observed. We also find that the distribution of centroid velocities is remarkably similar between clumps, despite spanning more than one order of magnitude in mass. We propose that our results are best explained by the mixing of gravitationally-driven multi-directional inflows, resulting in highly complex and dynamic hub centres. We also discuss the implications of the observed differentiated filament and hub gas kinematics in the context of the 3D morphology of hub-filament systems.
\end{abstract}

\begin{keywords}
stars:formation -- stars:massive -- ISM:clouds -- methods:observational -- submillimeter:ISM -- techniques:interferometric
\end{keywords}



\section{Introduction}
\label{sec:introduction}
Hub-filament systems (HFSs) are small networks of filaments that converge towards a higher density region where star formation activity is observed \citep{Myers2009}. The most luminous ($L \ge \qty{e5}{\Lsol}$) young stellar objects within the Galaxy are systematically embedded within HFSs \citep{Kumar2020,Peretto2022}. However, HFSs are not unique to high-mass star formation, but can also be found towards low-mass star-forming regions \citep[e.g.][]{Myers2009,Kirk2013}, making them structures of global interest when it comes to star formation theories. As a result, in the past 15 years or so, a large number of different scenarios have been proposed to explain their formation. Most of them start with a compression phase whereby a sheet of gas is formed first, soon followed by filament formation \citep[e.g][]{Nakamura2008,Myers2009,Wang2010,Gomez2014,Balfour2015,Inoue2018, Padoan2020}. The origin and associated physical scale of the compression differ according to the models, along with the physics that lead to the formation of filaments. The convergence of those filaments is either linked to the formation of the sheet itself \citep{Myers2009}, to the result of gravity that drags filaments towards the bottom of the potential well \citep[e.g.][]{Li2006,Vazquez-Semadeni2019}, to the consequence of multiple turbulence-driven sheet-sheet collisions \citep[e.g.][]{Inutsuka2015,Federrath2016, Padoan2016} or filament-filament interactions \citep{Kashiwagi2023,Kashiwagi2026}, or to the impact of oblique shocks on magnetised clouds \citep{Nozaki2026}. In those models, star formation naturally occurs at the convergence point of the filaments, i.e. the hub, since gas density is the largest there. One of the key differences between those models is the physical connection between the filaments and central hub, with, in particular, one fundamental question that still needs answering: Do filaments regulate the mass growth of their central hub and the cores within it?

In parallel to the theoretical effort mentioned above, numerous observational studies have investigated the density structure and gas kinematics of hub-filament systems \citep[e.g.][]{Liu2012,Peretto2013,Kirk2013,Busquet2013,Peretto2014,Dewangan2017,Williams2018,Trevino-Morales2019,Arzoumanian2021,Zhou2022,Liu2023,Rigby2024,Hacar2025}. Interestingly, the vast majority of those studies interpret their data in the context of collapse scenarios, whereby filaments feed the central hub via gravity-driven mass inflows. Evidence for such scenarios is based on the presence of velocity gradients along filaments, infall signatures in some cases, and the mass segregation of cores within the hub. However, only very few studies have the necessary angular resolution, spatial coverage, and sensitivity to follow the gas flows along individual filaments and down to the hub. For instance, with more than 100 clumps observed with ALMA at $\sim2\si{\arcsecond}$ resolution, the ATOMS project \citep{Liu2020} statistically investigated the kinematics of HFSs \citep{Zhou2022,Xu2023}. However, by focusing on the hub, the connection with the surrounding filaments is only partially addressed. Most of the other studies have mapped entire HFSs at similar angular resolution, but targeting only single sources that are often selected for their ability to form high-mass stars \citep[e.g.][]{Peretto2013,Beltran2022}, preventing then to draw general conclusions on the physical link between filaments and hubs.

In \cite{Anderson2021} (hereafter Paper I) we presented an analysis of the core population of a sample of 6 infrared dark hub-filament systems mapped with ALMA. We showed that a higher fraction of a clump's mass ends up within the most massive core (MMC) of IR-dark hubs compared to a sample of IR-bright clumps taken from \cite{Csengeri2017}. Here, we investigate the dense gas kinematics of those same 6 HFSs in order to characterise the dynamical link between the filaments and the hub, and eventually with the mass of the most massive core.

The details of the combined ALMA  total power (TP), 7\si{~\meter}, and 12\si{~\meter} observations are described in \secref{sec:observations}. In \secref{sec:nthp_j_1_0}, we calculate the integrated intensity to column density conversion factor for \nthp. We present our newly developed fully automated, multiple velocity component, hyperfine line fitting code \mwydyn{} in \secref{sec:model_fitting} (along with a detailed description in \appref{app:mwydyn}), and present the results obtained when applied to our \nthp(J=1--0) data of the six HFSs. In \secref{sec:discussion}, we discuss the 3D morphology of HFSs, the dynamics and energy balance along the transition from filament to hub and it relation to the mass of the most massive core within them. Finally, we present our summary and conclusions in \secref{sec:conclusions}.

\section{Observations}
\label{sec:observations}

\subsection{Sample selection}
\label{sub:sample_selection}
The sample studied here is the same sample of 6 HFS as presented in \citetalias{Anderson2021}. All 6 objects are infrared-dark clouds selected from the \cite{Peretto2009} \emph{Spitzer} dark cloud catalogue with a well defined hub-filament morphology seen in 8\si{~\micron} extinction. All lie at a distance between $\sim$2\si{~\kilo\parsec}--3.2\si{~\kilo\parsec}, have a peak column density $N_\mathrm{H_2}>10^{23}\si{\per\cubic\cm}$, and cover a broad range of masses from $\sim$300\si{~\Msol}--5000\si{~\Msol} (see \tabref{tab:obs_prop} and  \tabref{tab:sample}).

\begin{figure*}
	\centering
	\includegraphics[width=0.95\textwidth]{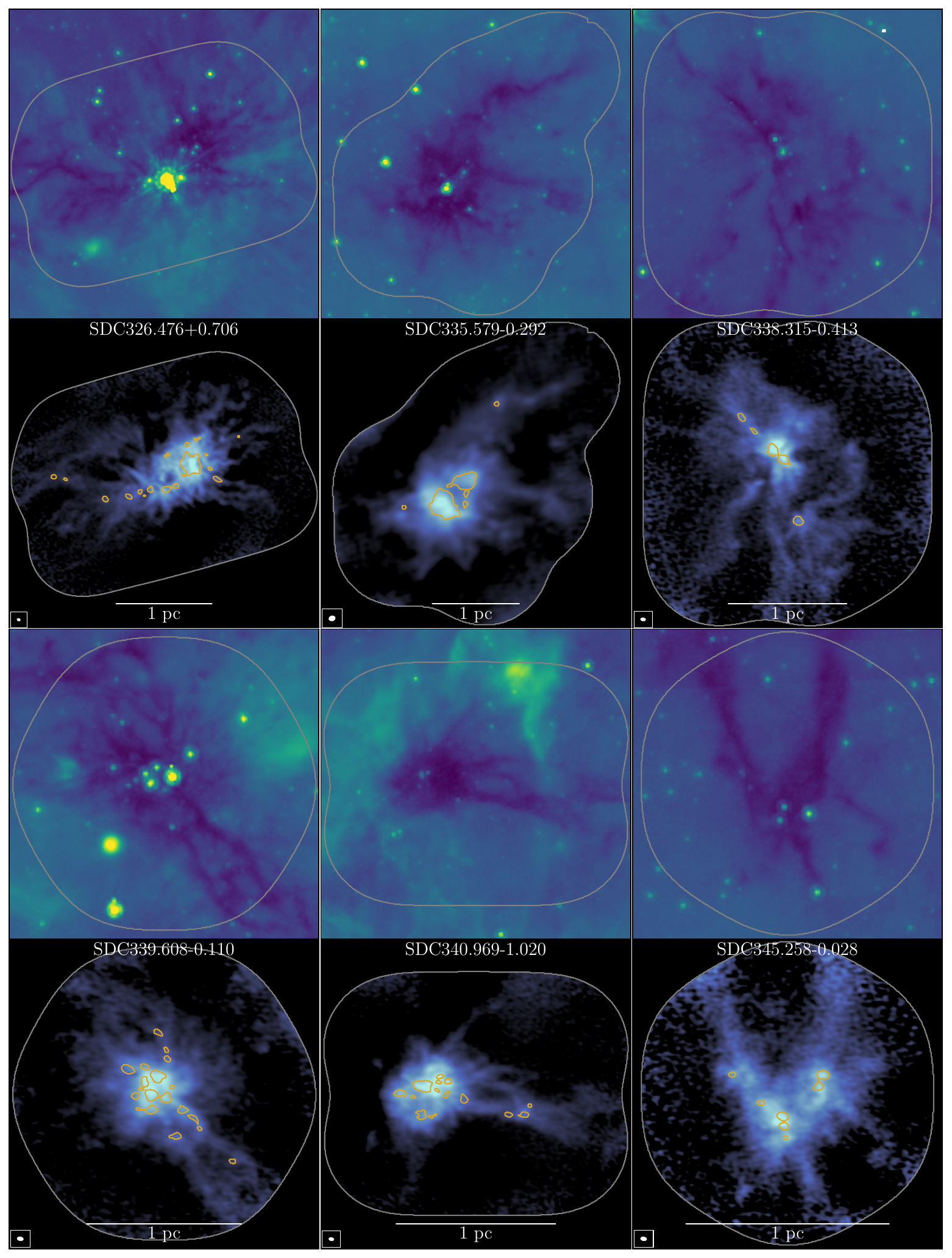}
		\caption{\emph{First and third rows}: \emph{Spitzer} 8\si{~\micron} images of the six IRDCs we observed with ALMA, showing prominent extinction features in a hub-filament system configuration. Below each Spitzer image is the corresponding ALMA combined 12\si{~\m}+7\si{~\m}+TP \nthp(1--0) integrated intensity images. The synthesised beam size of each image is shown in the lower left corner, the grey contour shows the extent of our ALMA fields. The orange contours denote the footprints of the 2.9\si{~\milli\m} continuum cores \citep{Anderson2021}.
		}
	\label{fig:ALMA-Spitzer_fields}
\end{figure*}

\subsection{ALMA observations}
\label{sub:alma_observations}
The observations we use here are part of the same observing ALMA cycle 3 programme\footnote{With the exception of cycle 0 data used for SDC335} as  described in \citetalias{Anderson2021}, although that paper focused on the continuum data, and the current paper focuses on the \nthp(J=1--0) dataset. We therefore refer the reader to \citetalias{Anderson2021} for a detailed description of the observations. In the following, we provide all additional information regarding the \nthp(J=1--0) dataset.

\input{obs_stats.tex}

The bands were centred around the frequency of the brightest hyperfine transition of \nthp(J=1--0) at \qty{93173.7643}{\mega\hertz}, with a spectral resolution of \qty{0.2}{\kilo\meter\per\second}. In addition to the \qty{12}{\meter} and \qty{7}{\meter} array observations, single-dish observations were taken of all fields using the \qty{12}{\meter} TP telescopes that are part of the ACA. These were taken in order to complement the combined \qty{7}{\meter}+\qty{12}{\meter} spectral line setups by providing the zero-spacing information, and hence recover the large scale emission in the fields. The total amount of TP observations for all six fields was $\sim \qty{36.3}{\hour}$. The TP data were combined with the combined \qty{7}{\meter}+\qty{12}{\meter} data using the CASA task \texttt{feather()}, with $\mathrm{sdfactor}=1$.

	\hspace{-3cm}
\begin{figure*}
	\includegraphics[width=\textwidth]{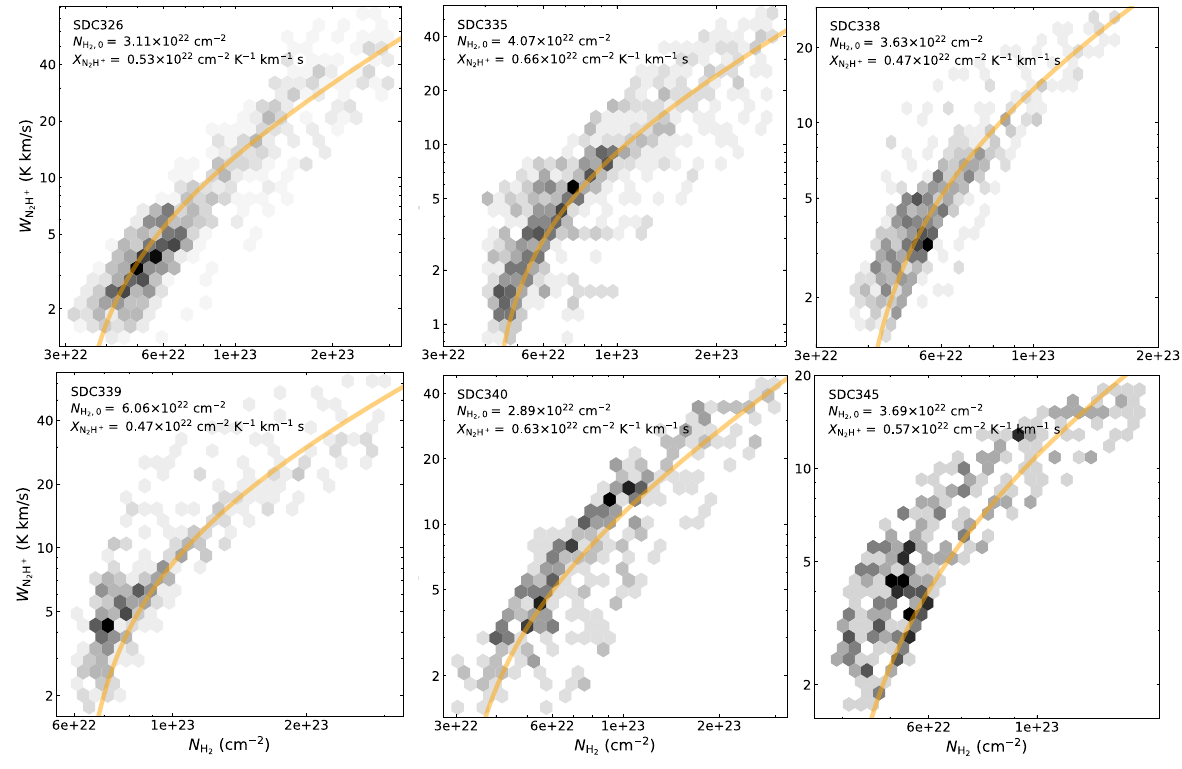}
	\caption{Histogram of the integrated intensity of \nthp{} against column density derived from \emph{Herschel} for each cloud. The integrated intensity maps were convolved and regridded to match the \emph{Herschel} column density map resolution. The column density offset, $N_\mathrm{H_2,0}$, is the column density at which we do not detect any \nthp{} emission in our maps. The orange line represents our best linear fit, see Eq.~(\ref{eq:hr_coldens}).
	} 
	\label{fig:nthp_xfactor}
\end{figure*}

For the final, fully combined \nthp(J=1--0) datacubes we achieve an angular resolution of around $\qtyrange{3.3}{5.2}{\arcsecond}$, which at the distance of the targets corresponds to a linear resolution of $\sim \qtyrange{0.033}{0.082}{\parsec}$, with a mean RMS noise (in line-free channels) between $\qtyrange{0.08}{0.18}{\kelvin}$. \tabref{tab:obs_prop} contains a summary of the observational parameters for the six fields.\footnote{As can be seen in \tabref{tab:obs_prop}, SDC335 is a bit of an outlier in terms of its observational properties. This is a result of the \qty{12}{\meter} component of the data for SDC335, which was taken during Cycle 0 with only 16 antennas in a less extended array configuration than the later Cycle 3 \qty{12}{\meter} data for the other five fields.}

\subsection{{\it Spitzer} and {\it Herschel} data}
\label{sub:spitzer_wise_and_herschel_data}
We use H$_2$ column density maps at a resolution of $\sim18\si{\arcsecond}$ which were constructed using the 160\si{~\micron} and 250\si{~\micron} data from the \emph{Herschel} Hi-GAL survey \citep{Molinari2010}, following the method presented in \cite{Peretto2016}.  For the construction of these maps we assumed a dust-to-gas mass ratio of 1\% and a specific dust opacity law $\kappa_{\lambda}=10\left(\frac{\lambda}{300\mu\rm{m}}\right)^{-1.8}$cm$^2$g$^{-1}$. We also use publicly available \emph{Spitzer} GLIMPSE 8\si{~\micron} data\footnote{\url{https://irsa.ipac.caltech.edu/data/SPITZER/GLIMPSE}} \citep{Churchwell2009} at an angular resolution of $\sim2.4\si{\arcsecond}$. \figref{fig:ALMA-Spitzer_fields} shows the \emph{Spitzer} 8\si{~\micron} fields for all six IRDCs, along with the ALMA \nthp(1--0) integrated intensity fields.

\section{\nthp(J=1--0) integrated intensity maps and column densities}
\label{sec:nthp_j_1_0}

In this paper, we focus on the analysis of the \nthp(J=1--0) emission line. With a critical density of $\sim \qtyrange{3e4}{6e4}{\per\cc}$ \citep{Shirley2015}\footnote{For a temperature of \qtyrange{50}{10}{\kelvin}, respectively.}, and a formation pathway that prevails in low CO abundance environments (i.e. where CO is depleted), \nthp{} preferentially traces cold, dense gas \citep{Caselli1995,Caselli2002,Bergin2002}. \nthp(J=1--0) integrated intensities have been shown to correlate extremely well with \molh{} column densities estimated from dust emission for the $\sim \qtyrange{e22}{e23}{\per\sc}$ regime \citep{Tafalla2001,Andre2007,Hacar2018,Barnes2020,Peretto2023,Priestley2023,Priestley2023a,Tafalla2023}. Towards higher column density cores, the agreement between dust emission and \nthp(J=1--0) worsens \citep{Peretto2013}, either because of opacity or the destruction of \nthp{} through the heating of dust grains and the release of CO due to protostellar activity \citep{Sternberg1995,Bergin2002,Bergin2007}. 

Another interesting aspect of the \nthp(J=1--0) line is that it has a hyperfine structure that can (if the velocity dispersion of the gas is low enough) be resolved into 7 components, some of which blend together in higher velocity dispersion environments into 3 groups of lines, one of which can sometimes be isolated. As a result of the existence of those hyperfine components at fixed spectral separation, line fitting must be performed in order to provide accurate measurements of the gas systemic velocity and velocity dispersion\footnote{See \appref{app:comparison_of_moments_vs_fitting} for a comparison of the FWHM obtained from fitting hyperfine structure vs calculating the 2\textsuperscript{nd} moment.}, and also estimates of the opacity of the line.

\figref{fig:ALMA-Spitzer_fields} shows the \emph{Spitzer} \qty{8}{\micron} images of the 6 infrared dark HFSs along with the corresponding \nthp(J=1--0) integrated intensity maps. The footprints of the cores identified in \citetalias{Anderson2021} are displayed as orange contours. The visual comparison between the filamentary extinction features and the integrated emission clearly shows that \nthp(J=1--0) is an excellent tracer of the infrared dark regions. We also notice that the outlines of the cores identified in dust continuum do not appear to obviously match clear overdensities in the \nthp(J=1--0) maps, despite the presence of \nthp(J=1--0) emission towards the vast majority of core footprints.

In order to better quantify the correlation between \molh{} column densities and the \nthp(J=1--0) integrated intensities, we constructed a pixel-by-pixel density plot of those two quantities. For that purpose, we computed \molh{} column density maps from \emph{Herschel} at  $18\si{\arcsecond}$ resolutipom (see Section \ref{sub:spitzer_wise_and_herschel_data}) for each of the 6 sources.  We then convolved our \nthp(J=1--0) integrated intensity maps to the same angular resolution as the column density maps, and also matched their astrometry and pixel scales. The corresponding density plots of each source are shown in \figref{fig:nthp_xfactor}. One can see that, despite the presence of some scatter, both quantities are reasonably well correlated across the board. We further quantify those correlations by performing a linear fit using the following model: 
\begin{equation}
	\label{eq:hr_coldens}
	N_\mathrm{H_2}  = N_\mathrm{H_2,0} + X_\mathrm{N_2H^+} W_\mathrm{N_2H^+}
\end{equation}
where $N_\mathrm{H_2}$ is the \emph{Herschel} column density,  $N_\mathrm{H_2,0}$ is the median column density where the $\mathrm{SNR}\sim1$ in our integrated integrated maps\footnote{Note that the six clumps we are analysing here sit in different regions of the Galactic plane and thus have very different structural backgrounds. Differences in $N_\mathrm{H_2,0}$ values are a reflection of those different backgrounds.}, $X_\mathrm{N_2H^+}$ is the intensity to column density conversion factor, and $W_\mathrm{N_2H^+}$ is the \nthp(J=1--0) integrated intensity.

Here, the only free parameter is the conversion factor $X_\mathrm{N_2H^+}$. The best fit and associated best fit parameters for each cloud are displayed in \figref{fig:nthp_xfactor}. We notice that the conversion factor $X_\mathrm{N_2H^+}$ has very similar values for each of the 6 fields, all within the range $X_\mathrm{N_2H^+}=(0.55\pm0.12)\times10^{22}\unit{\per\sc \per\kelvin \per\kilo\meter \second}$. This value is very similar to the value obtained by \cite{Hacar2018} for the Integral Shape Filament (ISF) in Orion, for a temperature of $\sim \qty{12}{\kelvin}$. 

New clump masses were derived using these column density conversion factors, which are shown in \tabref{tab:sample}, alongside the masses derived from \emph{Herschel} column density contours presented in \citetalias{Anderson2021}. The differences between these masses is discussed in \secref{sub:energy_budget}.

\input{sample_stats.tex}

\section{\nthp(J=1--0) model fitting}
\label{sec:model_fitting}
Across the six HFSs, over \num[group-separator={,}]{180000} spectra with $\mathrm{SNR}>10$ are detected, amongst which a significant fraction displays evidence of multiple velocity components. In order to extract the gas kinematics from those spectra, one needs to fit them in an automated way with little to no supervision in order to be both reproducible and unbiased.

\begin{figure*}
	\centering
	\includegraphics[width=0.95\textwidth]{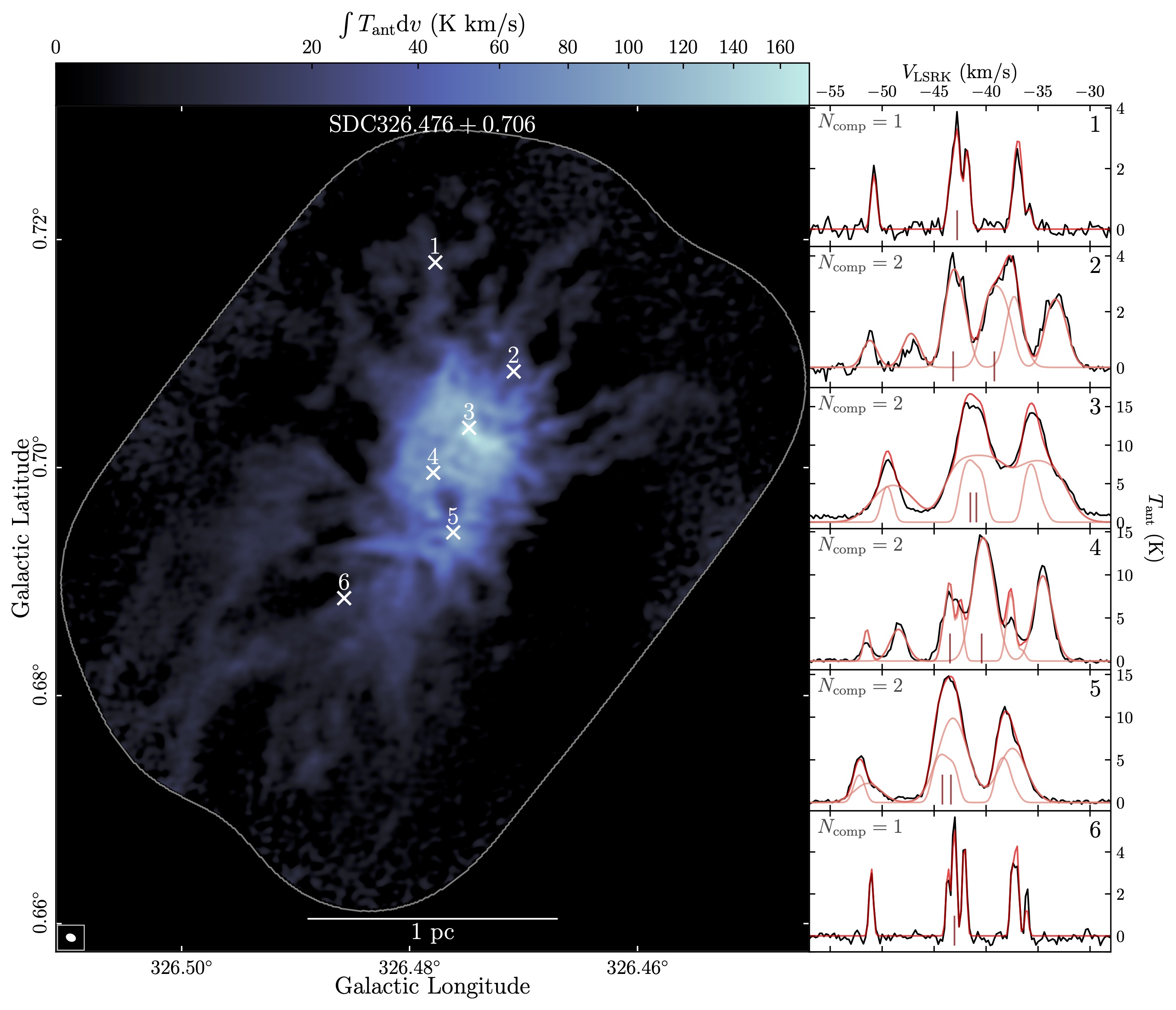}
		\caption{Integrated intensity \nthp{} image of SDC326, with example spectra in black (numbered 1--6, with locations marked on the map), and their corresponding best fitting models produced by \mwydyn{} in red. For models comprised of multiple velocity components, their constituent sub-models are shown in light red. The locations of the best fit velocity centroids are marked with dark red vertical bars.
		}
	\label{fig:SDC326_example}
\end{figure*}

While several semi-/fully-automated multiple Gaussian component fitting algorithms are available  (e.g., \textsc{SCOUSE} - \citealp{Henshaw2016}; \textsc{GaussPy+} - \citealp{Riener2019}; \textsc{BTS} - \citealp{Clarke2018}) only a few exist for hyperfine line fitting (e.g. \textsc{PySpecKit} - \citealp{Ginsburg2022}; \textsc{HfS} -  \citealp{Estalella2017}; MUFASA - \citealp{Chen2020}, McFine - \citealp{Williams2024}).  In this work, we use \mwydyn{} (\emph{Welsh}: worm; \emph{pronounced}: muy-din; IPA: \textipa{["mU\textsubarch{i}dIn]}), a fast and fully automated multiple velocity component hyperfine fitting code, originally developed specifically for the analysis of this dataset but now fully adapted for more generalised use, and available on GitHub.\footnote{\url{https://github.com/mphanderson/mwydyn}} \mwydyn{}  has already been successfully used by \citet{Rigby2024}, and a detailed description of how the code works can be found in \appref{app:mwydyn} with examples of fits shown in  \figref{fig:SDC326_example}. 

In this paper, we fitted all \nthp(1-0) spectra with a peak signal-to-noise ratio (SNR) $\ge10$, which, in the optically thin case, corresponds to a peak SNR on the weakest isolated component $\ge 4.3$.

\subsection{Global distributions}
\label{sub:global_distributions}
\begin{figure*}
	\centering
	\includegraphics[width=0.95\textwidth]{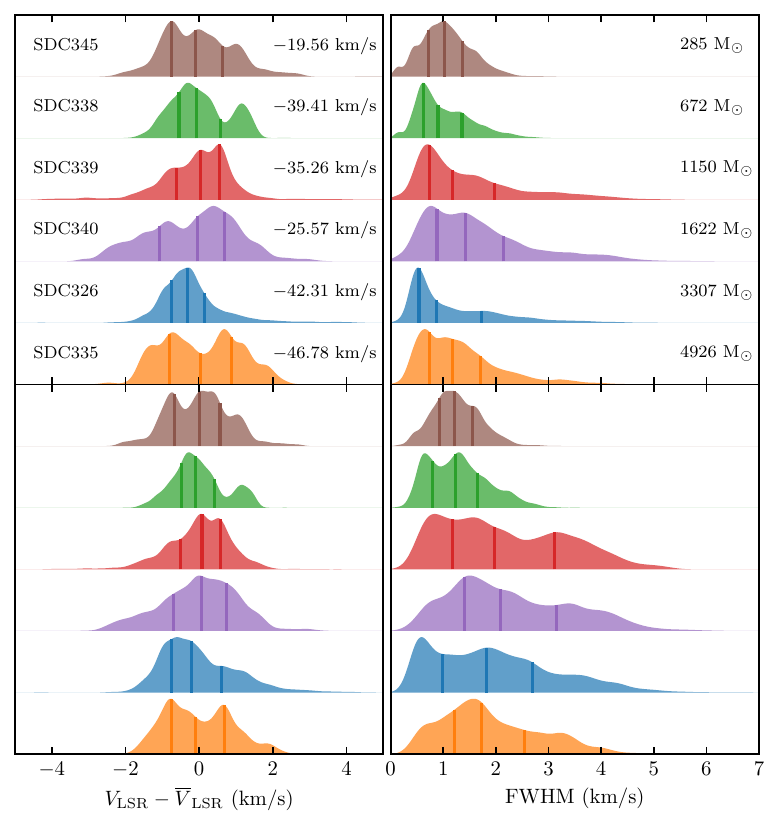}
	\caption{\emph{(top row)} Distributions of fitted velocity centroids (left) and fitted FWHM (right) produced by \mwydyn{}. \emph{(bottom row)} The same as above, but with each value weighted by the integrated intensity of their respective submodel. The integrated intensity-weighted mean centroid velocity ($\overline{V}_\mathrm{LSR}$) of each cloud (shown to the right of each distribution) has been subtracted from their respective velocity distributions to aid comparison. The distributions have been ordered in ascending order of clump mass derived from \nthp, shown in the top-right panel. Quartiles are marked with vertical lines.
	}
	\label{fig:global_distributions}
\end{figure*}

Here we present the global distributions of the \mwydyn{} fit properties for the six hub-filament systems in our sample. The two fit properties we focus on are the fitted velocity centroids (\vlsr{})\footnote{Velocities presented in this paper were calculated with respect to the rest frequency of  the brightest hyperfine component. See \tabref{tab:multiplet}.} and the linewidths (FWHM). The global distributions of both of these quantities are shown for each hub-filament system in \figref{fig:global_distributions}, where the distributions have been ordered by clump mass. Note that these are the distributions of all fitted velocity components. The two upper panels are the raw kernel density estimations (KDEs) of fitted properties, and the two lower panels have been weighted by the integrated intensity of each component. We present these weighted distributions as they should be more representative of the amount of mass in the system that has a given value of a quantity, given that the integrated intensity correlates reasonably well with the column density (see \figref{fig:nthp_xfactor}). For example there are considerably more spectra in the outer filamentary regions of the clumps compared to the hub region (which also have less intensity, and likely less mass), so the unweighted distributions are biased by the relative area of the different parts of the clump.

The \vlsr{} distributions are, in broad terms, similar between hubs. All of the distributions cover a similar range in velocity of $\sim \qtyrange{3}{4}{\kilo\meter\per\second}$, and there does not seem to be any correlation with the mass of each clump. This appears to be the case in both the raw and intensity-weighted distributions. All of the distributions display small peaks that are separated by roughly \qty{1}{\kilo\meter\per\second}, which may in part be due to the filamentary substructures we see in the data. These peaks are somewhat less apparent in the intensity-weighted distributions. It is clear that based on the global distribution of velocities alone we would not be able to tell which clump is the most massive, contains the highest mass core, or has the highest $f_\mathrm{MMC}$ , i.e. the fraction of the clump mass that sits within the most massive core (see \tabref{tab:sample}).

The FWHM distributions, shown in the right column of \figref{fig:global_distributions}, are more varied. Beginning with the unweighted distributions we see that the higher FWHM tail of the distributions is far less extended when looking at the two low mass hubs compared to the higher mass hubs in the sample. One similarity in all of the distributions (except SDC345) is a large peak at a FWHM value of $\sim0.6\si{~\kilo\meter\per\second}$, accompanied by a strong shoulder at around 1--2\si{~\kilo\meter\per\second}. This implies that the vast majority of spectra (in terms of number) are dominated by narrow linewidth, low dispersion gas, close to the isothermal sound speed. 

When weighting the FWHM distributions by the integrated intensity of each fitted submodel, the multimodality of the distributions become far more apparent, with what was a shoulder in the unweighted distribution appearing more like an additional peak in the distribution. There is also significantly more power in the FWHM wings, compared to the non-weighted distributions, indicating that the brighter (i.e. higher column density) regions of the cloud generally have a larger linewidth. Overall the location of the mean shifts towards higher FWHM with increasing clump mass, and in particular, the proportion of the distribution above a given FWHM seems to increase with clump mass. This behaviour occurs up to a mass of around $\qty{e3}{\Msol}$, and then decreases slightly. However, this sample size is far too small to make any conclusions regarding this trend.

The lowest mass clump (SDC345) has a peak at a FWHM of around \qty{1}{\kilo\meter\per\second} (in the unweighted distributions), in comparison to the roughly \qty{0.6}{\kilo\meter\per\second} peak of the other clumps. This is somewhat unexpected given how IR-dark it is, and the relative low-mass of the filaments that comprise the system. That being said the mapped area of SDC345 is limited to the densest central (hub) region, with very little coverage of the outer filamentary structures seen in the \emph{Spitzer} data. This, combined with the overall lower SNR of the \nthp{} data for this field (probably due to the lower column density of the object) means that only a small, central patch (with $\mathrm{SNR}>10$) of the mapped area was fit by \mwydyn{}. We speculate that the lack of a lower linewidth peak may be due to the limited mapping of the outermost filamentary structures, or that the energy involved in the formation of the hub has not been dissipated yet.

Throughout the remainder of the paper our analysis will focus on the integrated intensity-weighted fit properties unless otherwise stated.

\subsection{Spatial distributions}
\label{sub:spatial_distributions}

\begin{figure*}
	\centering
	\includegraphics[width=\textwidth]{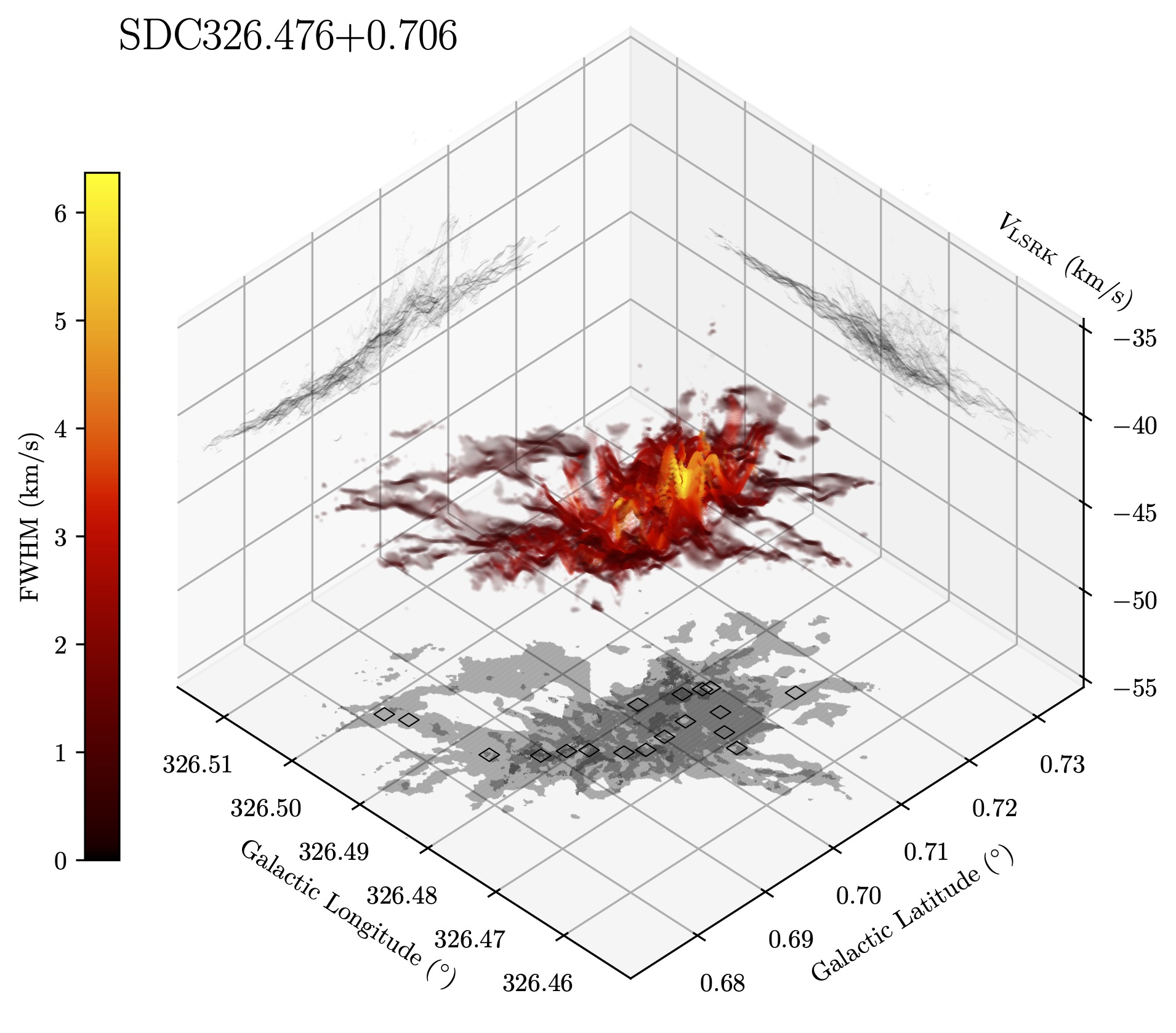}
	\caption{3D PPV Galactic longitude-latitude-\vlsr{} ($l$, $b$, $v$) visualisation of the fit results produced by \mwydyn{} for SDC326. The points are coloured by the best fit FWHM value. On the lower box surface is a 2D projection ($l$, $b$) showing the number of fitted velocity components along the line of sight, with darker grey indicating more components. The positions of the ``cores'' presented in \protect\cite{Anderson2021} are marked with boxes. The left and right surfaces show 2D projections in both ($b$, $v$) and ($l$, $v$), respectively.
	}\label{fig:SDC326_PPV_3D}
\end{figure*}

Here we examine how the properties of the gas vary spatially across each of the hub-filament systems. In \figref{fig:SDC326_PPV_3D} we show a 3D visualisation of our fit results for SDC326 in position-position-velocity (PPV), i.e. Galactic longitude-latitude-\vlsr{} ($l$, $b$, $v$), with the points coloured by the fitted FWHM. On each of the three surfaces in the plots are 2D projections along the axis perpendicular to the given surface. The bottom surface represents the number of fitted components along the line of sight, with the darkest grey representing 3-component fits, and the lightest grey representing 1-component fits. On the two side panels are position-velocity (PV) projections of the fit results (Galactic latitude-velocity and longitude-velocity on the left and right, respectively). The darker grey represents a high number of points along that projected axis. The visualisations for all of the hubs are shown in \figref{fig:PPV_3D}.

Despite the varied appearance and mass of the six hubs, we see many consistent attributes in their PPV visualisations. All of the structures appear to be coherent in velocity, with filamentary features seen in all hubs. We see that the gas has typically low velocity dispersion in the outskirts, which gradually increases towards the hub centres. The number of fitted components also increases towards the hub centres (e.g. \figref{fig:SDC326_example}), with some increases elsewhere that may be due to overlapping filamentary structures or the presence of cores. 

In the PV projections, it also appears that the dispersion in the centroid velocities increases toward the centres. Note that the points plotted here are the raw values only, and have not had their appearance modified/weighted by the integrated intensity of their respective submodel. Some of the clumps (SDC339, SDC340) exhibit quite clear clump-scale velocity gradients, with only slight gradients present in SDC335 and SDC345. No clear velocity gradients are seen for SDC326 and SDC338 at the clump scale. 

Small, sinusoidal features are seen in PV in all of the objects. Similar features have been seen in other work and often interpreted a signature of gravity-driven flows \citep[e.g.][]{Henshaw2016,Henshaw2020,Rigby2024,Alvarez-Gutierez2024,Sandoval-Garrido2025}. It is important to note that these features are not an artefact of the fitting process, as these are clearly visible within PV diagrams shown in \figref{fig:SDC326_pv_diagram} and \figref{fig:pv_diagrams}. Although a full characterisation of the filament population will be the topic of a follow-up paper, we can already comment on the fact that a large fraction of those velocity oscillations are not associated with cores\footnote{At least, cores detected within our continuum data.}. The origin of these oscillatory features remains thus unclear. 

\subsection{Radial trends}
\label{sub:radial_trends}
\begin{figure}
\vspace{-0.2cm}
	\centering
	\includegraphics[width=0.9\columnwidth]{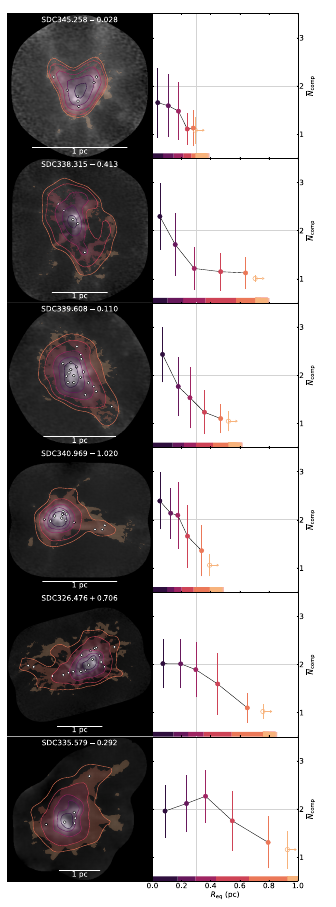}
	\vspace{-0.3cm}
		\caption{\emph{Left}: \nthp{} integrated intensity images for each cloud, with \emph{Herschel} column density contours superimposed. The contour values, in units of $10^{22}$~cm$^{-2}$, are [6.0, 6.5, 8.3, 11.1, 15.0]  for SDC345, [5.0, 5.6, 7.3, 10.1, 14.0] for SDC338, [7.5, 9.0, 13.6, 21.3, 32.0]  for SDC339, [8.0, 14.4, 33.5, 65.4, 110.0]  for SDC340, [5.5, 8.9, 19.1, 36.2, 60.0] for SDC326, and [6.0, 9.7, 20.8, 39.2, 65.0] for SDC335. Core locations are marked with diamonds. \emph{Right}: Mean number of fitted components within each contour level as a function of $R_\mathrm{eq}$. Vertical lines represent the standard deviation of $N_\mathrm{comp}$. Unfilled circles represent the outermost points for which we do not have an $R_\mathrm{eq}$ value, and hence their radii are lower limits.
		}
	\label{fig:ncomp_radius}
\end{figure}

In the previous section we explored some of the spatial features and trends in a qualitative manner. Here, we focus on investigating how the gas properties behave as a function of radius more quantitatively. In \figref{fig:ncomp_radius} and \figref{fig:kde_radius} we divide our hub-filament systems into H$_2$ column density bins using  contours on the {\it Herschel} column density images, and examine how the distributions of fit properties vary within each annulus. While those bins were selected to  cover approximately similar areas, the outermost contour of each clump was set at a value that encompasses the vast majority of the \nthp{} emission that was fit with \mwydyn{}. As a result of different  background column densities, those values vary from clump-to-clump (from $5\times10^{22}$~cm$^{-2}$  for SDC338 to $8\times10^{22}$~cm$^{-2}$ for SDC340).  The  spacing between contours were then derived using a square-root scaling. This was used in part because the column density drops off as $\sim1/r^2$, resulting in contours with equivalent radii ($R_\mathrm{eq}$) across clumps. Points that do not lie within the outermost contour level are grouped together into one distribution, and hence do not have an $R_\mathrm{eq}$ value assigned to them, i.e. the $R_\mathrm{eq}$ of the outermost contour acts as a lower limit on their radius value. 

First we will consider the mean number of fitted components, i.e. $\overline{N}_\mathrm{comp}$ (see \figref{fig:ncomp_radius}). As seen in the PPV visualisations in \figref{fig:PPV_3D} we see that the number of fitted components increases with decreasing radius. This is largely an expected behaviour given the increased complexity of the spectra that we see towards the centre (e.g. as in \figref{fig:SDC326_example}) that require a more complex model (i.e. more components). When comparing these plots between HFSs, we see that the increase in $N_\mathrm{comp}$ occurs at larger radii in the higher mass systems. Within $R_\mathrm{eq} < \qty{0.3}{\parsec}$, $N_\mathrm{comp}$ increases steadily from SDC345 (the lowest mass) to SDC335 (the highest mass), with the region of multiple-component fits extending in radius. Those regions with extended complex kinematics are also some with more extended high-column density regions, suggesting that higher column densities are achieved via dynamic gas inflows. We also notice that  for the two most massive HFSs, this increase in number of components appears to flatten off at around \qtyrange{0.2}{0.4}{\parsec}. For SDC335 we see a decrease at smallest radii. This is driven by two factors: i) a ring of spectra around the third contour that show prominent self absorption, which \mwydyn{} attempts to fit with higher-complexity models; ii) The spectra in the centremost region are actually rather simple, with one peak, a very broad linewidth and wings, hence generally needing fewer components to model. See \figref{fig:SDC335_example} for a set of example spectra and models for SDC335. 

In all cases, \figref{fig:kde_radius} (3rd column) shows that the FWHM distributions peak at low values (i.e. $<\qty{1}{\kilo\meter\per\second}$) at large radii, and then increases steadily towards the central contour level. For SDC345, the distributions are fairly similar across all radii, which again may be due to the limited mapping as described in \secref{sub:global_distributions}. The number of peaks within the distributions also vary steadily with increasing mass until SDC326, with a more subtle change in the distributions for SDC335. Note that the spatial resolution is $\sim1.6$ times worse for SDC335 that the rest of the sample, so we are probably resolving less substructure within it.

The velocity centroid distributions (\figref{fig:kde_radius} - 2nd column)  overall show a smaller degree of variation (in terms of breadth) with radius, and the variations with radius are less consistent across clouds. In some cases (e.g. SDC335) have trends that are the inverse of trends in the FWHM distributions, i.e. the distributions get simpler and narrower with decreasing radius. These distributions are complex, and it is difficult to extract more meaningful conclusions from them as opposed to the FWHM distributions, especially as the shapes of these distributions will be heavily influenced by the 3D morphologies of the HFSs and also their 2D projections on the sky. 

What is clear from those distributions of velocity dispersions and centroid velocities is that they both carry important information regarding the intrinsic kinematics and energetics of the HFSs.  While the fitted FWHM of each velocity component gives us an idea of the velocity dispersions along the line of sight, to quantify the total kinetic energy of the entire system we also need to include the component-to-component contribution, which we can estimate as the velocity difference between each of the fitted velocity components and the centroid velocity of the system as a whole. In order to obtain a more global picture, we combined them in the form of a single total velocity dispersion measurement $\sigma_\mathrm{kin}$ (for each pixel), defined as:
\begin{equation}
	\label{eq:sigkin}
	\sigma_\mathrm{kin} = \sqrt{
	\frac{\sum_i^{N_\mathrm{comp}} w_i [(v_i - \bar{v})^2 + \sigma_i^2]}
	{\sum_i^{N_\mathrm{comp}} w_i}}
\end{equation}
where the sum is over the number of fitted components $N_\mathrm{comp}$ in a given spectrum, $v_i$ and $\sigma_i$ are the fitted velocity centroid and velocity dispersions of the $i$-th component\footnote{Note that we do not explicitly fit for $\sigma_i$ in \mwydyn{}, so this is simply converted from the FWHM by: $\sigma_i=\mathrm{FWHM}_i/2\sqrt{2\ln2}$.}, $\bar{v}$ is the mean velocity of the HSF, estimated as the intensity-weighted mean of all fitted centroid velocities in a given cloud. The weight, $w_i$, in our calculation is the integrated intensity of the $i$-th component.

\begin{figure}
	\centering
	\vspace{-3cm}
	\includegraphics[width=\columnwidth]{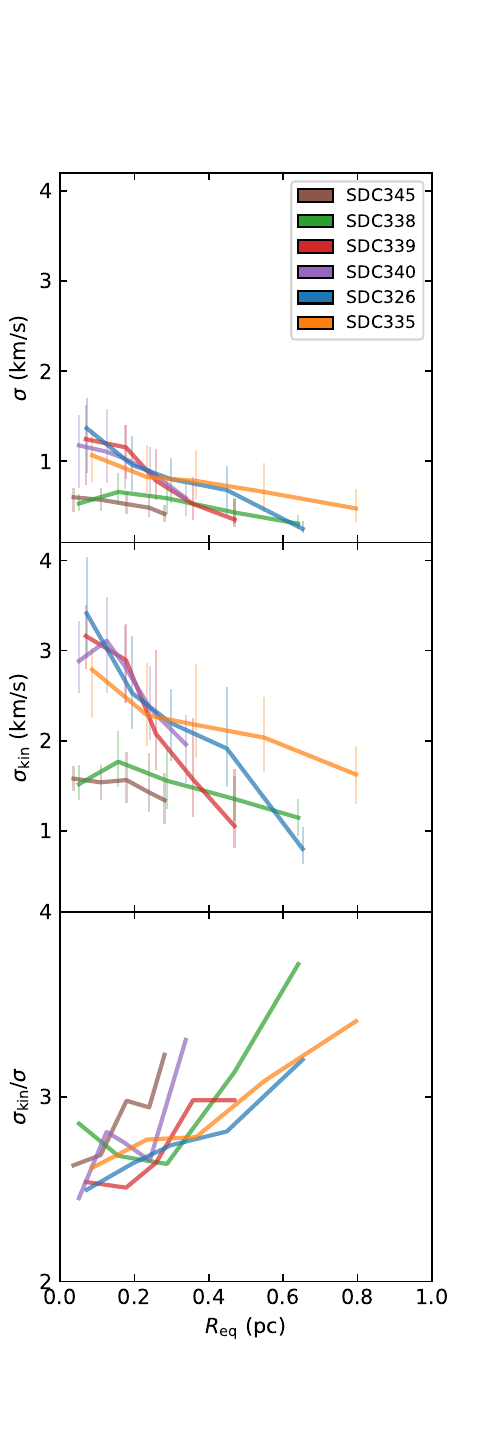}
	\vspace{-2cm}
	\caption{Comparison across HFSs of the median velocity dispersions (top), and median total velocity dispersions (middle) calculated for each radial bin. The errorbars represent the interquartile range of the distributions shown in \figref{fig:kde_radius}. The bottom panel shows the ratio $\sigma_{\mathrm{kin}}$ to $\sigma$ median profiles. 
		}
	\label{fig:sig-sigkin_radius}
\end{figure}

By comparing $\sigma$ and $\sigma_\mathrm{kin}$ one is able to estimate the relative contribution of the individual line-of-sight velocity dispersion ($\sigma$) versus the contribution of the inter-component variations in the velocity centroids to the kinematics. This is similar to what \citet{Traficante2020} have done albeit on much larger scales. \figref{fig:sig-sigkin_radius} displays such a comparison as a function of $R_\mathrm{eq}$ (the full distributions of $\sigma_\mathrm{kin}$ as a function of radius are shown in \figref{fig:kde_radius}). We can see that, at large radii, where the filaments lie, the total velocity dispersion is dominated by the inter-component kinematics, with $\sigma_\mathrm{kin}$ is around 3 to 4 times larger than $\sigma$. At the smallest radii, this ratio drops to around 2 to 3. We also notice that the two clumps with the smallest masses (SDC345 and SDC338) display flatter velocity dispersion profiles and a much more similar relative contribution of $\sigma$ to $\sigma_\mathrm{kin}$ at all radii. This difference is most likely due to the lower mass concentration (i.e. density) of those two clumps compared to the other ones (see \figref{fig:mass_radius}), leading to weaker gravitational acceleration towards their centres.

\section{Discussion}
\label{sec:discussion}

\subsection{The 3D morphology of IR-dark HFSs: spheres vs sheets}
\label{sub:3d_morphology}
Many models for the formation of HFSs start with the formation of a compressed sheet of gas. However, finding observational evidence of the existence of those sheets is extremely difficult, mostly because of the observations' inability to probe the depth of a cloud along the line-of-sight. Using observations of different transitions of the same molecule one can  estimate its average volume density which, when combined with a \molh{} column density map, can provide the line-of-sight depth of a cloud. In the few cases that this has been done, the filamentary nature of the clouds have been confirmed \citep{Li2012,Bonne2020}. However in the case of Musca, a \qty{6}{\parsec} long filamentary cloud located around \qty{140}{\parsec} from the Sun, it has been argued that its global morphology is rather that of a sheet-like cloud seen edge-on \citep{Tritsis2018,Tritsis2022}. Here, we discuss the 3D morphology of HFSs through the lens of their dense gas kinematics.

There is an ever growing body of evidence that parsec-scale star-forming clumps are in a state of global collapse, whether those clumps are hub-filament systems \citep{Peretto2013,Kirk2013}, or not \citep{Peretto2006,Schneider2010,Ragan2015,Traficante2018,Traficante2023}. In fact, \citep{Peretto2023} showed that most infrared dark clumps are dynamically decoupled from their parent molecular clouds as a result of clump collapse. It is therefore most likely that the six hub-filament systems targeted within our study are no exception, and are also collapsing. One of the sources in this sample (SDC335), was studied by \cite{Peretto2013} and \cite{Xie2026}, has clear evidence for being in a state of global collapse. 

\begin{figure}
	\centering
	\includegraphics[width=0.95\columnwidth]{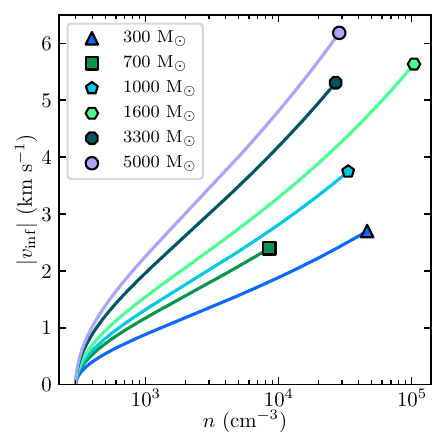}\\
	\includegraphics[width=0.95\columnwidth]{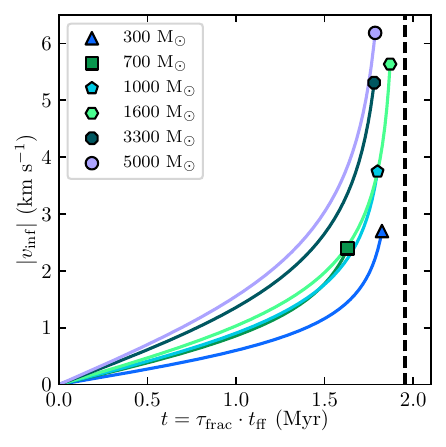}
		\caption{Infall velocity evolution as a function of density (top) and time (bottom)  for six spherically symmetric collapsing clumps with masses [$300, 700, 1000, 1600, 3300, 5000$]~\si{\Msol} and final radii [$0.3, 0.7, 0.5, 0.4, 0.8, 0.9$]~\si{\parsec}, respectively. These values were chosen to be comparable with the six HFSs that our within our sample. All clumps are assumed to have the same initial density of $n_0=\qty{3e-2}{\per\cc}$. $\tau_\mathrm{frac}$ is a time variable represented as a fraction of the initial free-fall time at initial density $n_0$ (see \appref{app:clump_infall_velocity}). The vertical dashed line shows the initial free-fall time.
		}
	\label{fig:vinf}
\end{figure}

When considering the velocity distributions in \figref{fig:global_distributions} (bottom left), the strong similarity of the centroid velocities between HFSs is challenging to explain if one considers a collapsing, fragmented, spheroidal clump. Indeed, gravitational acceleration being larger for the more massive/denser clumps, SDC335 ($M\sim\qty{5000}{\Msol}$) should show evidence of more dynamic gas kinematics than SDC345 that is $\sim20$ times less massive. In order to quantify the differences in expected acceleration between the different clumps of our sample we computed the instantaneous free-fall velocities $v_\mathrm{inf}$ of uniform density spheres of initial density $n_0=\qty{300}{\per\cc}$, a typical molecular cloud volume density \citep{Peretto2023} and initial radii $r_0=r(n/n_0)^{1/3}$, where $r$ and $n$ are the observed external clump radii and densities. Assuming that the mass o the collapsing clump is constant, we are able to derive  $v_\mathrm{inf}(t)$ and $v_\mathrm{inf}(n)$ ( \citealp{Girichidis2014} - see also \appref{app:clump_infall_velocity} for their derivation). \figref{fig:vinf} shows that the expected free-fall velocities for our clump sample vary between $\sim\qty{2}{\kilo\meter\per\second}$ for the least massive ones to $\sim\qty{6}{\kilo\meter\per\second}$ for the most massive. If the filaments that connect to the hub were to be randomly distributed within free-falling spheres, then a similar range of centroid velocity distribution widths should be observed. As evidenced by  the very similar range of centroid velocities  observed in all clumps (see \figref{fig:global_distributions} bottom left), this is not the case. 

Another possible 3D configuration for our sample of infrared dark HFSs is, as expected from a number of models (see \secref{sec:introduction}), a sheet. However, if collapsing sheets were to be randomly oriented with respect to the plane-of-the-sky we would also expect large differences in the centroid velocity distributions of clumps of different masses. Therefore, the only viable collapsing sheet scenario is one where infrared dark HFSs are sheet-like clumps seen mostly face-on. While at first it could seem unlikely to have selected, by chance, only face-on sheet-like clumps, our source selection is heavily biased. Indeed, by design, we selected sources that exhibit clear hub-filament morphologies in mid-infrared extinction,  an observational bias that might favour the selection of face-on  sheet-like clumps.  However, the extrapolation that all HFSs (including the infrared bright ones) are face-on sheets seems, given the increasing number of HFSs being identified, unlikely.

When discussing the 3D morphology of HFSs, another element has to be considered. \figref{fig:global_distributions} (bottom right) shows that, unlike the centroid velocities, the FWHM (and hence velocity dispersion) does correlate with mass, i.e. broader distributions for higher mass clumps, across  a range that is not dissimilar to that expected from free-fall velocities (see above). A similar trend has been identified by \cite{Rigby2024}. However, those large velocity dispersions are not randomly spread over the clumps, but are almost exclusively located within the central hub regions (see \figref{fig:sig-sigkin_radius} and \figref{fig:kde_radius}). In a sphere-like configuration, this behaviour could be explained by the fact that streamlines are aligned with the line-of-sight when observing the hub centre leading to a maximum velocity dispersion of the gas, and perpendicular to it when observing the outskirts of the HFS. In a sheet-like configuration, this significant increase of the velocity dispersion towards the hub could be the consequence of the dynamical interaction of the gas flows channelled by the multi-directional filaments (but confined to a sheet) produces a randomisation of the flow directions in the hub region, leading to larger line-of-sight velocity dispersions. However, infall signature towards SDC335 using HCO\textsuperscript{+}(J=1--0) was observed, therefore leaving no doubt that at least this particular HFS is collapsing on pc-scale along the line-of-sight \citep{Peretto2013,Xie2026}.

 The main limitations of the sheet vs sphere discussion presented above are linked to the small sample size we are dealing with, and the lack of clear and unique observational signatures of the formation and evolution of HFSs in various models. Both aspects will be soon tackled within the upcoming LAGAF (PI: N. Peretto) and Panta Rei  (PIs: N.Peretto, A.Traficante, S. Clarke, M. Merello) ALMA programmes.

\begin{figure*}
\centering
	\includegraphics[width=\textwidth]{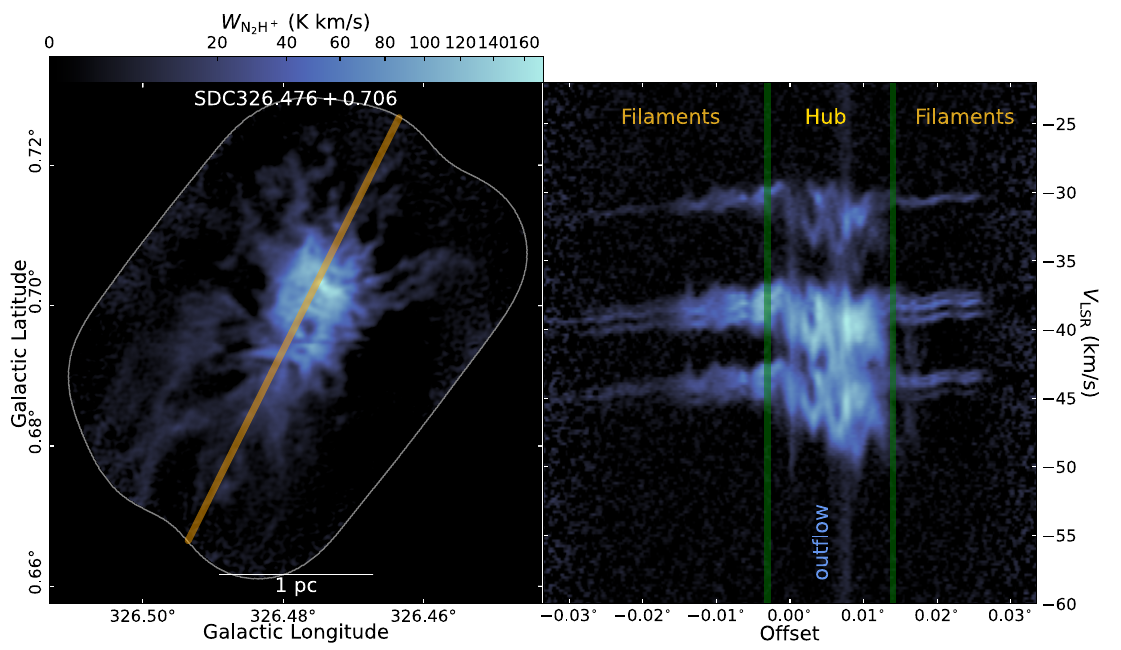}
	\caption{The left-hand panels show the integrated \nthp(J=1--0) intensity data for SDC326, with the orange line indicating the axis for which the PV-diagram was generated, which is shown in the right-hand panel. The vertical green lines on the right-hand-side plot approximately indicate the transition between the filament and hub regions. }
	\label{fig:SDC326_pv_diagram}
\end{figure*}

\subsection{The filament-hub transition}
\label{sub:filament_hub_transition}

\figref{fig:SDC326_pv_diagram} shows a position-velocity (PV) diagram generated from an example cut across the hub-filament system SDC326. PV diagrams for all hubs are shown in \figref{fig:pv_diagrams}. All six HFSs show  a net dynamical differentiation between the outer filament-dominated and hub-dominated regions. How sharp the transition between the two regions is depends on the HFS, with the most notable (in PV) found towards SDC326. While looking at a different problem, i.e the formation of brown dwarfs, \cite{Bonnell2008} looked at the kinematics of hub-filament like structures within their star-cluster formation simulations\footnote{See their Figure 4.}. The similarities in terms of density and velocity structures are apparent, even though on different scales. In that study, they propose that after the formation of the first few massive stars at the bottom of the gravitational potential well, gas velocity dispersion there is large as it gets virialised, while inflowing gas through filaments is being gently accelerated at low velocity dispersion. It is the internal dynamics of the cluster-hosting hub that generates a clear dynamical differentiation from the rest of the clump.%

Another example of a numerical simulation that studied clump dynamics is provided by \cite{Lee2016}, where a flattened hub is forming at the centre of a globally collapsing cloud. The hub itself is stabilised as the result of its rotational and turbulent energy while accreting mass from its surrounding cloud. Such a scenario also leads to a clear dynamical differentiation between the outer filamentary region and the hubs.

In both examples provided above, low velocity dispersion flows of gas are being accreted onto a highly dynamic, high density structure. This type of configuration could naturally lead to a low velocity shock at the outer boundary of the growing hub. How such shocks would appear in \nthp(J=1--0) observations is not clear. However the relatively sharp transitions that appear in at least some of the PPV visualisations and PV diagrams (where the filaments meet the hub centre), might be evidence of such a process.

Note that, relative to the sheet vs sphere discussion (see \secref{sub:3d_morphology}), sharp discontinuities such as that observed in SDC326 might only be expected for sheet-like HFSs, a consequence of the dynamical differentiation of the virialised hub and the inflowing filaments. A more smooth transition (as seen for most of the other HFSs - see \figref{fig:pv_diagrams}) is expected in the spherical case where varying projection angles from centre to edge is responsible for the differentiation. It might be therefore be the case that different HFSs can be formed from varying initial conditions and 3D morphologies.

Finally, \figref{fig:SDC326_pv_diagram} shows evidence for the presence of outflows. The fact that they are identified in \nthp(J=1--0) is a clear indication that they could be responsible, at least to some extent, for increasing the dense gas velocity dispersion within the hubs. In the case of SDC335, the identification of outflows and the characterisation of their properties have suggested that there is a direct link between the outflowing mass rate and the mass infall rate of the parent HFS \citep{Avison2021}. A full characterisation of the protostellar outflow population within all six HFSs and their impact on the hub kinematics will be the topic of a forthcoming paper.

\subsection{The energy budget of hub-filament systems}
\label{sub:energy_budget}
In order to determine the energy budget of the six HFSs we constructed their virial ratio profiles, i.e. the ratio of the kinetic energy over gravitational energy. To do this, we first need to derive the mass profiles of each HFS. This is done by first obtaining the \nthp(J=1--0) integrated intensity within the same contours as those used in \figref{fig:ncomp_radius} and then convert them into mass by using the $X_\mathrm{N_2H^+}$ factors derived in \secref{sec:nthp_j_1_0}. Note that the total masses derived this way differ from the values obtained in \citetalias{Anderson2021} by a factor of up to 3 in the case of SDC338 (see \tabref{tab:sample}). These differences come from the two methods that have been used to derive masses. In \citetalias{Anderson2021} clump masses correspond to the bijective masses \citep{Rosolowsky2008} enclosed within a column density contour of $3 \times \qty{e22}{\per\sc}$ in the \cite{Peretto2016} column density maps whose large-scale structures had been removed with a \qty{10}{\arcminute} median filter. This contour was chosen as a compromise so that we had reasonable boundaries for all 35 clumps studied within that paper, however in reality some of the cloud structures seen in \emph{Spitzer} \qty{8}{\micron} extinction and \nthp{} emission are not contained within said contours, hence probably underestimating their masses. In the present study, masses are directly estimated from the \nthp{}  emission line (from $X_\mathrm{N_2H^+}W_\mathrm{N_2H^+}$), guaranteeing the selection of the entirety of the high density gas within each clump. These differences, combined with the imperfect correlation between \molh{} column density and \nthp(J=1--0) integrated intensity, probably account for most of the discrepancies. 

\begin{figure}
	\centering
	\vspace{-0.8cm}
	\includegraphics[width=\columnwidth]{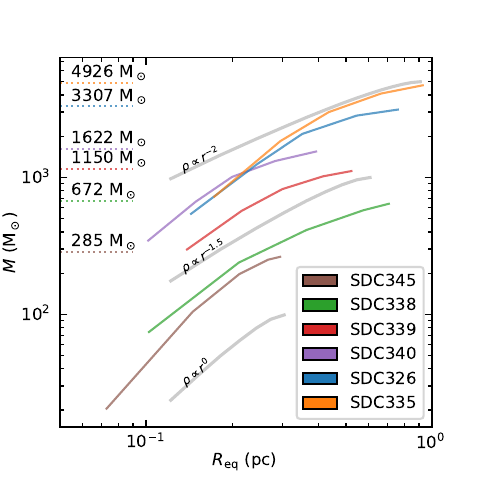}
		\caption{Enclosed mass within each contour level (see \figref{fig:ncomp_radius}), as a function of the equivalent radius ($R_\mathrm{eq}$) of that contour level. The dotted line represents the total mass from our \nthp(J=1--0)-derived column density maps, including the points which lie outside of the outermost column density contour, and hence have an undefined $R_\mathrm{eq}$. The three grey profiles correspond to three single density power models derived following the method described in \citet{Peretto2023} }
	\label{fig:mass_radius}
\end{figure}

\figref{fig:mass_radius} presents the mass profiles of the six HFSs. We first notice that the profile shapes are very similar to one another, and that while on a log-log plot, they are not straight lines. For spherical clouds, it has been shown that even for power-law mass profiles of the form $M \propto R_\mathrm{eq}^{\alpha}$, curved profiles are expected as a result of projection effects \citep{Peretto2023}. This is demonstrated by the three spherical model profiles shown in \figref{fig:mass_radius}. However, for face-on sheets, those projection effects are minimised and the change in the slope of the mass profiles directly links to a change of the mass surface density profile slope. The profiles here are broadly consistent with $M \propto R_\mathrm{eq}$ in the filament region and $M \propto R_\mathrm{eq}^2$ in the hub region, which, in terms of radial dependence of the mass surface density $\Sigma$ translates into $\Sigma \propto R_\mathrm{eq}^{-1}$ and $\Sigma \propto R_\mathrm{eq}^{0}$ (i.e. constant), respectively. However, we have to stress that these \nthp-based mass profiles are very uncertain, in particular within the hub region where systematics (i.e. line optical depth, excitation temperature variations, abundance variations) could affect the shape of those profiles. For instance, a constant mass surface density profile within the central $\sim \qty{0.8}{\parsec}$ region of SDC335, as suggested by \figref{fig:mass_radius}, should be spatially resolved in the \emph{Herschel} column density map as a central flat plateau, something that we do not observe (see column density contours in \figref{fig:ncomp_radius}).

Combining the estimated mass profiles with the velocity dispersion profiles presented in \figref{fig:sig-sigkin_radius}, we can derive their virial ratio profiles. The virial ratio $\alpha_\mathrm{vir}$ is often used as a tool to discuss whether a cloud is bound or unbound, but here we merely use it as a measure of the balance between kinetic and gravitational energy. The virial ratio is defined as:
\begin{equation}
	\label{eq:vir_par}
	\alpha_\mathrm{vir} = \frac{2 E_\mathrm{kin}}{|E_\mathrm{grav}|}
\end{equation}
where the kinetic energy $E_\mathrm{kin}$ of an enclosed mass $M$ with 1D velocity dispersion $\sigma_\mathrm{1D}$ is defined as:
\begin{equation}
	\label{eq:ekin}
	E_\mathrm{kin} = \frac{3}{2}M \sigma^2_\mathrm{1D}
\end{equation}
and the gravitational potential energy $E_\mathrm{grav}$ of an enclosed mass $M$ with radius $R$ is given by:
\begin{equation}
	\label{eq:egrav}
	E_\mathrm{grav} = -\frac{(3-k_\rho)}{(5-2k_\rho)} \frac{G M^2}{R}
\end{equation}
where $G$ is the gravitational constant, and $k_\rho$ is the power law index of the density profile of the clump. 

For an oblate clump, the total (i.e. estimated at the outermost radius) gravitational energy is very similar to the spherical case \citep{Bertoldi1992}. However, the radial profiles of the gravitational potential of oblate clumps are different to those of spherical clumps \citep[e.g.][]{Lee2016a}, and so the mass located outside a given radius $R_\mathrm{eq}$ does impact the calculation of the gravitational potential energy. However, some recent numerical work have shown that the interaction between the mass within the clump and the gravitational potential external to it does not contribute much to its total gravitational energy \citep{Ganguly2022}, even though it might artificially lead to systematics (i.e. higher virial ratios) towards cores \citep{Ballesteros-Paredes2018}.

\begin{figure}
	\centering
	\includegraphics[width=\columnwidth]{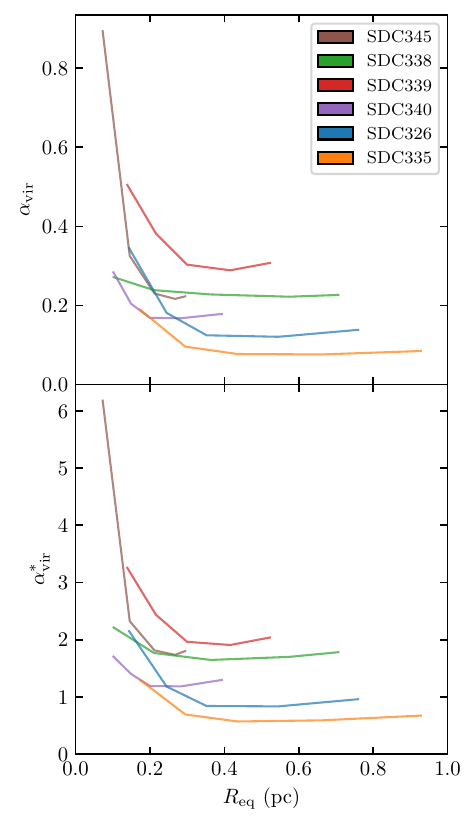}
	\caption{Virial parameter (of each contour level), as a function of the equivalent radius ($R_\mathrm{eq}$) of the outer boundary of that contour level. The upper panel shows the virial parameter calculated using the velocity dispersion $\sigma$ only ($\alpha_\mathrm{vir}$), with the lower panel using $\sigma_\mathrm{kin}$ ($\alpha_\mathrm{vir}^*$).
		}
	\label{fig:virial_radius}
\end{figure}

\figref{fig:virial_radius} shows the virial ratio profiles obtained with $k_{\rho}=2$ \citep{Williams2000,Peretto2023}, for both velocity dispersion profiles (i.e. $\sigma(R_\mathrm{eq})$ and $\sigma_\mathrm{kin}(R_\mathrm{eq})$). While the exact values of the virial ratios are uncertain (see above), all HFSs show a similar behaviour whereby the virial ratio increases from values typical of gravitationally bound structures (i.e. $\alpha_\mathrm{vir} \lesssim 2$) within the filament region to values typical of unbound structures (i.e. $\alpha_\mathrm{vir} \gtrsim 2$) towards the hub region. The central increase of the virial ratio could be due to a number of different factors. First, it could be due to the presence of protostellar outflows injecting kinetic energy most efficiently within the inner part of  the hub, leading to an increase of the velocity dispersion there. Also, it could be due to an underestimate of the clump mass at small radii due the poorer correlation between H$_2$ column density and integrated \nthp(J=1--0) intensity at high densities/temperatures. Finally,  \cite{Ballesteros-Paredes2018} have argued that higher virial ratios towards cores can be attributed to ignoring the surrounding mass in the gravitational energy calculation. Together, high virial ratio values in the densest regions of HFSs are to be expected.

\subsection{Determining factors for the mass of the most massive cores}
\label{sub:mmc_correlations}
In \citetalias{Anderson2021} we showed that, on average, the fraction of mass in a hub-filament system that was locked up its most massive core $f_\mathrm{MMC}$ is larger for infrared dark HFSs compared to infrared-bright clumps. These fractions could reach values as high as $\sim24\%$ for SDC335. Here, we want to understand what physical characteristics of HFSs might set those high values $f_\mathrm{MMC}$, as they are central to our understanding of high-mass star formation.

\begin{figure}
	\centering
	\includegraphics[width=\columnwidth]{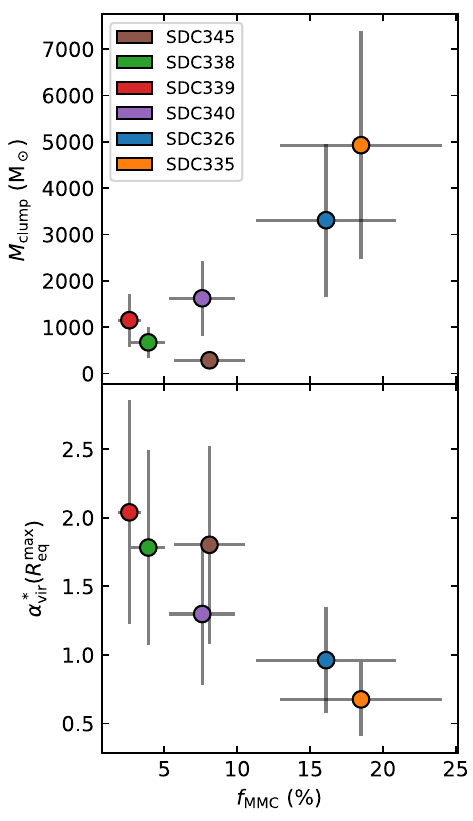}
	\caption{Clump mass and virial parameter at the outermost clump contour against the $f_\mathrm{MMC}$ value for each hub. Errors of 10\%, 20\%, and 30\% on the velocity dispersions, clumps distances, and $X_\mathrm{N_2H^+}$ values, respectively, have been propagated to obtain the errorbars shown on this plot.  
		}
	\label{fig:fmmc_plots}
\end{figure}

First, as mentioned in \secref{sec:nthp_j_1_0}, the masses of the HFSs have been revised. While clump masses published in \citetalias{Anderson2021} were derived within a common column density contour, the masses published here are derived from \nthp(1-0) bright gas, which is equivalent to measuring mass above a common volume density \citep{Priestley2023}.  We have therefore used those new masses to calculate new $f_\mathrm{MMC}$ estimates (see \tabref{tab:sample}). The changes are most significant for the two lower-mass HFSs, SDC345 and SDC338\footnote{The factor of 3 change in the mass of SDC338 is primarily a result of the  increase in the clump boundary area compared to the one defined in \citetalias{Anderson2021}, where the clump area was around 2.5 times smaller.}. In \figref{fig:fmmc_plots}, we show a scatter plots of $f_\mathrm{MMC}$ versus clump mass and virial ratio of the outermost enclosing contour. One can see that, while there is a correlation with clump mass, there is a tentative anti-correlation with $\alpha_\mathrm{vir}$. Even though we are aware that this is a very small sample, this could be a suggestion that the global dynamical state of the clump is important in determining the mass of the most massive cores within clumps. This is reminiscent of the findings in \cite{Bonnell2011}, who showed that only clumps with low virial ratios will develop to form high-mass stars. We note that the clump with the highest global $\alpha_\mathrm{vir}$ value, SDC339, has both a large number of cores (19), yet no core with a mass greater than $M_\mathrm{core} \sim \qty{30}{\Msol}$, and that the most massive core in the sample, with $M_\mathrm{core} \sim \qty{900}{\Msol}$, belongs to the hub with the lowest global virial ratio (SDC335). Interestingly, such correlations were not found by \cite{Rigby2024} based on their sample of 7 IRDCs. The reason behind this discrepancy could be linked to the HFS nature of our sample which might be more efficient at concentrating their mass within a single core, and that this efficiency also correlates with  the mass of the clump and its virial ratio. However, the low number statistics of both studies prevent us from drawing robust conclusions. The anti-correlation seen in \figref{fig:fmmc_plots} could also be an indication of the rapid evolution of accreting clumps \citep{Traficante2023} whereby the fractional mass growth of  the most massive cores is larger than that of their  parent collapsing clumps. In that scenario, SDC335 and SDC326 would be more evolved versions of SDC338 and SDC339, and their observed larger luminosities \citep{Anderson2021} could be taken as evidence of this.

\section{Summary and conclusions}
\label{sec:conclusions}
We presented new ALMA observations of a sample of six infrared dark hub-filament systems. We analysed the dense gas kinematics of those sources through the fitting of the \nthp(J=1--0) emission line using the new fully automated, multiple velocity component, hyperfine structure fitting code \mwydyn. 

This analysis shows that the hub and filament regions of hub-filaments systems are dynamically distinct, i.e. filaments are made of quiescent gas with low-velocity dispersion, while hubs present extremely complex kinematics alongside very large velocity dispersions. Also, the distribution of velocity centroids across this sample of HFSs are extremely similar between clumps, despite spanning a factor of $\sim 20$ in mass. In some HFSs the velocity profile across the filament-hub transition is relatively sharp, a feature that might be consistent with accretion shocks at the outer boundary of the hub. We discussed these different features in the context of the 3D morphology of hub-filament systems, comparing and contrasting expectations from free-falling sheets and spheres. Interestingly, none of those two configurations completely match the observables for all HFSs, possibly indicating that both configurations might be valid depending on the target. However, projection effects, combined with uncertain 3D density and velocity fields, make it hard to infer what the different collapse signatures might be between a HFS within a sheet or within a sphere.  This clearly highlights the need for synthetic observations of \nthp(J=1--0) of collapsing structures to further investigate the complex gas dynamics within the centres of HFSs.

Finally, we found a tentative anti-correlation between $f_\mathrm{MMC}$ and the global $\alpha_\mathrm{vir}$ value for the six HFSs, suggesting that the HFS global virial ratio might set what fraction of its mass will end in the most massive core. The fact that such correlation was not found in earlier clump studies might point either towards the specific ability of HFSs to focus the mass within one massive core, or possibly towards the limitations of small number statistics.

Characterising the dense gas velocity fields within clumps and their relation to the formation and evolution of  cores is the key to answer the question of how massive stars gain their mass. This study demonstrates that we have now both the data and the tools to unveil that link. However, a number of limitations are also clearly apparent, both in terms of setting up the expectations with synthetic observations of collapse simulations, and in terms of sample size and sample biases. It is only by tackling both sets of limitations that we will reach the next level in our understanding of the earliest stages of massive star formation.

\section*{Acknowledgements}
	MA is supported by the Science and Technology Facilities Council (STFC). NP and AJR acknowledge the support of the STFC consolidated grant number ST/S00033X/1. NP also acknowledges support from of the STFC Small Award grant number APP30146. SER acknowledges support from the consolidated grant (ST/K00926/1) from STFC. ADC acknowledges the support from the Royal Society University Research Fellowship (URF\textbackslash R1\textbackslash 191609 and URF\textbackslash R\textbackslash 241028). G.A.F. acknowledges support from Deutsche Forschungsgemeinschaft (DFG, German Research Foundation)  through SFB 1601 ‘Habitats of massive stars across cosmic time’ (sub-project B1) and under Germany’s Excellence Strategy EXC 3037 – 533607693– Our Dynamic Universe.  G.A.F. also acknowledges support from the University of Cologne and its Global Faculty programme. The UK ALMA Regional Centre (ARC) Node is supported by STFC grant numbers ST/Y004108/1 and ST/T001488/1. GMW acknowledges support from STFC under grant number ST/R000905/1. This paper makes use of the following ALMA data: ADS/JAO.ALMA\#2011.0.00474.S,\\ ADS/JAO.ALMA\#2015.1.01014.S, and\\ ADS/JAO.ALMA\#2016.1.00810.S. ALMA is a partnership of ESO (representing its member states), NSF (USA) and NINS (Japan), together with NRC (Canada), MOST and ASIAA (Taiwan), and KASI (Republic of Korea), in cooperation with the Republic of Chile. The Joint ALMA Observatory is operated by ESO, AUI/NRAO and NAOJ. This research made use of the Python packages Astropy\footnote{\url{https://astropy.org/}} \citep{AstropyCollaboration2013,AstropyCollaboration2018,AstropyCollaboration2022}, IPython\footnote{\url{https://ipython.org/}} \citep{Perez2007}, LMFIT\footnote{\url{https://lmfit.github.io/lmfit-py/}} \citep{Newville2014,Newville2022}, Matplotlib\footnote{\url{https://matplotlib.org/}} \citep{Hunter2007}, NumPy\footnote{\url{https://numpy.org/}} \citep{Harris2020}, SciPy\footnote{\url{https://scipy.org/}} \citep{Virtanen2020}, and scikit-image\footnote{\url{https://scikit-image.org}} \citep{vanderWalt2014}. 
	This research also made use of NASA's Astrophysics Data System Bibliographic Services, analysisUtils \citep{Hunter2023}, CARTA\footnote{\url{https://cartavis.org}} \citep{Comrie2021},  TOPCAT\footnote{\url{http://www.star.bris.ac.uk/~mbt/topcat/}} \citep{Taylor2005}, and SAOImageDS9\footnote{\url{http://ds9.si.edu/}} \citep{Joye2003}.

\section*{Data Availability}

\emph{The data underlying this article are available in the article and in its online supplementary material. Any additional data will be shared on reasonable request to the corresponding author.}



\bibliographystyle{mnras}
\bibliography{hfs-paper-2} 



\appendix


\section{Description of \mwydyn}
\label{app:mwydyn}

At its core, our fitting programme is based upon the procedure and assumptions of the hyperfine structure fitting method from \textsc{CLASS}, a set of continuum and line emission analysis tools that is in turn part of the \textsc{GILDAS}\footnote{\url{https://www.iram.fr/IRAMFR/GILDAS}} software package \citep{Pety2018}. The hyperfine structure method fits a molecular line emission spectrum with an individual hyperfine multiplet with a model comprised of four free parameters: $T_\mathrm{main}\tau_\mathrm{main}$, $v_\mathrm{cen}$ (centroid velocity of the reference hyperfine component), $\Delta v$ (FWHM linewidth), and $\tau_\mathrm{main}$. We also refer to these parameters as $p_1$, $p_2$, $p_3$, and $p_4$, respectively. Here, $\tau_\mathrm{main}$ corresponds to the sum of the central opacities of each hyperfine component, and $T_\mathrm{main}$ to the sum of the central antenna temperatures of each hyperfine component in the optically thin case. The method depends on five assumptions:
\begin{enumerate}
	\item All hyperfine components have  the same excitation temperature
	\item The opacities of each hyperfine component have a Gaussian profile as a function of frequency
	\item Each hyperfine component has the same linewidth
	\item The multiplet components do not overlap
	\item The main beam temperature is well suited to the source (i.e. the source is well resolved)
\end{enumerate}

The total opacity (as a function of frequency/velocity) of the multiplet is expressed as:
\begin{equation}
	\label{eq:tau_v}
	\tau(v) = p_4 \sum^N_{i=1}r_i \cdot
	\exp\left[-4\ln2\left(\frac{v - \left(\delta v_i + p_2\right)}{p_3}\right)^2\right]
\end{equation}
where $p_2$ is the velocity of the reference component, $\delta v_i$ is the velocity offset of the $i\textsuperscript{th}$ component of the multiplet (relative to the reference component), $p_3$ is the FWHM linewidth, $r_i$ is the theoretical relative strength of each hyperfine component in the optically thin case (where $\sum_i^N r_i = 1$), $N$ is the number of hyperfine components in the multiplet, and $p_4$ is the sum of the opacities at the hyperfine component line-centres. 

The total line profile (which we use to fit our spectra) is then expressed as:
\begin{equation}
	\label{eq:t_ant_v}
	T_\mathrm{ant}(v) = \frac{p_1}{p_4} (1-e^{-\tau(v)})
\end{equation}
where $p_1$ is effectively the peak intensity of the line, scaled by the opacity.

For this study, we use the same model of the multiplet parameters for the \nthp(J=1--0) transition as used by \textsc{CLASS}, which are shown in \tabref{tab:multiplet}. We chose the brightest hyperfine component as the reference component for our model fitting.

\input{n2h+_model.tex}

Our fitting programme \mwydyn{} extends the \textsc{CLASS} hyperfine fitting procedure to be able to fit a superposition of multiple hyperfine multiplets to whole datacubes of spectra, as well as automating initial guess parameters and determining how many multiplets are required to produce a good fit. The algorithm also attempts to reduce the discontinuities between fit parameters between adjacent spectra. Here, we will outline all of the steps the programme runs though over the course of fitting an input data cube.
\begin{enumerate}
	\item A user specified configuration file read by  \mwydyn{} contains various options for the chosen hyperfine line model, fitting bounds, parallel processing settings, and whether to save the data products and produce summary figures.
	
	\item The input data cube is converted into units of brightness temperature (\si{~\kelvin}), from surface brightness (\si{~\Jy\per\beam}).
	
	\item The RMS of the first and last 25 channels are used to measure the noise level in all spectra in the cube, generating an RMS map. This is used to ensure that the algorithm only fits spectra that have a peak signal-to-noise ratio (SNR) greater than 10 (for the brightest hyperfine component). When $\tau \ll 1$, this corresponds to an $\mathrm{SNR}\sim4.3$ for the isolated component.
	
	\item All spectra that satisfy this SNR condition ($\mathrm{SNR_{lim}}=10$) are cycled through in succession. Initially one \nthp(J=1--0) hyperfine multiplet (hereafter velocity component) is fit to the spectrum. We use \textsc{lmfit} \citep{Newville2014}, a fitting package based around the Levenberg-Marquardt algorithm for our fitting. \textsc{lmfit} extends many of the methods from \textsc{SciPy}'s \texttt{optimize} module and provides a high-level interface for creating custom models, setting fit parameters guesses and  bounds, and also adds the ability to algebraically define relationships and constraints between model parameters. 
	
	We set out our initial guesses and bounds on the fit parameters as follows. For each pixel, \mwydyn{} locates the brightest channel in the spectrum, and use its amplitude as the the initial guess for $p_1$ and its velocity coordinate as the guess for $p_2$. The centroid velocity is allowed to vary $\pm \qty{20}{\kilo\meter\per\second}$ around this initial guess. The FWHM linewidth $p_3$ is initialised at $\qty{0.5}{\kilo\meter\per\second}$, which is roughly the isothermal sound speed at \qty{10}{\kelvin}, and is allowed vary between \qtyrange{0.1}{10}{\kilo\meter\per\second}. As \nthp(J=1--0) is generally considered to be optically thin, we set our initial guess of the total opacity $p_4$ as 0.2, with bounds between 0.1--30, consistent with the range adopted by \textsc{CLASS}. 
	
	In addition to the previously mentioned parameter bounds, we add a constraint that requires that the peak intensity of the total hyperfine multiplet is greater than $\mathrm{SNR_{lim}}$. This is less important for single velocity component fits, however when multiple velocity component models are fit this helps prevent adding numerous additional, very low intensity sub-models that would not be independently detected.
	
	\item After fitting a single velocity component model to the spectrum, the algorithm tries to fit a 2- and 3-component model to the spectrum. In principle the algorithm can fit as many velocity components as the use wishes, however with each additional sub-model another four free parameters are added, the parameter space becomes more complex and slower to explore, and hence the fitting takes significantly longer to converge. In this study, we imposed a limit of the maximum number of components of $N_\mathrm{comp}^\mathrm{max}=3$, leading to a reasonable computation time, and satisfactory fits in the most dynamically complex regions.
	
	For both the 2- and 3-component fits the initial guess for one of the submodels are set in the same manner as the 1-component model initial guesses. For the second and third set of four parameters (i.e. for the second and third velocity components/submodels), we copy the guesses and bounds for parameters $p_1$, $p_3$, and $p_4$, but adjust our guess for $p_2$. We estimate the velocity range of detected emission by finding the first and last channels where the emission is greater than $\mathrm{SNR_{lim}}/2 = 5$, meaning that there is less than a 1 in $\sim1.7\times10^6$ chance of a spurious noise spike in one channel registering as a false positive detection of emission. Once we have this range of detected emission, we compare this range to the range of velocities expected from a single hyperfine multiplet ($\sim15\si{~\kilo\meter\per\second}$). If the range of detected emission, for instance is $\sim18\si{~\kilo\meter\per\second}$ then we initialise the guesses for $p_2$ at $\pm3\si{~\kilo\meter\per\second}$, the sign depending on whether the range is shifted to lower or higher velocities compared to the range of the 1st fitted component.

	\item Once we have the set of 3 models for the current spectrum, the algorithm needs to determine the best fitting model to the data. We use the Bayesian Information Criterion (BIC) for evaluating the quality of the three models. Essentially, the BIC is the log-likelihood with an additional term that penalises the use of models with a larger number of free parameters, thus helping reduce the chance of overfitting. The model with the lowest BIC is chosen as the best model, but when additional velocity components are added (i.e. 2- and 3-component models) we require that there should be a significant improvement in the BIC in order to accept the larger N component fitting. In this study, we require  $\Delta\mathrm{BIC}=20$ to accept the more complex/higher-order model. 
	
	In all prior steps, each individual spectrum is fit independently of the surrounding spectra. Although the fitting algorithm explores the parameter space well and generally provides reasonable solutions, there may be some discontinuities in the fit parameters of adjacent pixels. Given that spectra that lie within the beam area are correlated, we do not expect strong discontinuities between the models in adjacent spectra. Therefore, we perform several iterations of identifying and refitting spectra whose models strongly deviate from their adjacent pixels, as follows:
	
	\item We cycle through all of our initial best fits for the spectra and identify model spectra within a radius of 2 pixels which have a lower BIC than the spectrum in question. If the spectrum within search radius had a lower BIC value, we refit the current spectrum by using the best fitting parameters from the  lower-BIC model spectrum  as initial guesses for the refit. If the new model spectrum has a BIC value that is significantly better (i.e. $\Delta\mathrm{BIC}=20$) than the previous model, we accept the new model as the best fitting model for the spectrum.
	
	\item We repeat this process 10 times across the cube to ensure that there is ample opportunity for the models to converge on a reasonable, locally consistent solution.
	
	\item With the fitting procedure now complete we write all of the results to a FITS table containing a list of all model/submodel parameters for all fitted spectra. We also write FITS cubes of the model spectra and residuals, and maps such as the the RMS, number of fitted components ($N_\mathrm{comp}$) in each spectrum, and the final model BIC values.
	
\end{enumerate}

Note that, as with CLASS, \mwydyn{} works for any hyperfine molecular line. \mwydyn{} supports user generated hyperfine models in the same format as our \nthp(J=1--0) model, but note that we have only extensively tested the programme with \nthp(J=1--0). Also, all of the fitting and refitting \mwydyn{} is fully parallelised, significantly shortening the runtime on multi-processor machines. The user can specify how many processors they wish to utilise for the programme.

\section{Comparison of moments vs. fitting}
\label{app:comparison_of_moments_vs_fitting}
\begin{figure}
	\centering
	\includegraphics[width=\columnwidth]{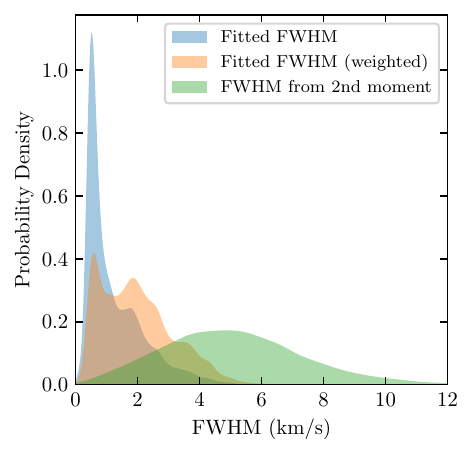}
		\caption{Comparison of the distributions of FWHM values for SDC326 obtained by fitting with \mwydyn{} (raw values in blue, and values weighted by the integrated intensity in orange), and by calculating the 2\textsuperscript{nd} moment of the isolated component (in green).
		}
	\label{fig:SDC326_FWHM_comparison}
\end{figure}

To illustrate the necessity to fit the full hyperfine structure of \nthp, \figref{fig:SDC326_FWHM_comparison} shows the kernel density estimations (KDEs) of FWHM values obtained from calculating the 2\textsuperscript{nd} moment of the isolated component, and fitted FWHM obtained using \mwydyn. It is immediately apparent that one would obtain a completely different picture of the kinematics when using the moment-based method, especially in objects similar to those in this sample that contain multiple velocity components along the line of sight, with complex blending.

\section{Clump infall velocity}
\label{app:clump_infall_velocity}

Here, we present estimates of the infall velocity for a set of clumps that are assumed to have formed from the free-fall collapse of a larger molecular cloud structure, using the analytical model outlined by \cite{Girichidis2014}. The free-fall time, $t_\mathrm{ff}$, is the time it takes for a completely static spherically symmetric parcel of gas (with density $\rho_0$) to collapse into a single point under gravity without any opposing forces, and is frequently used as a characteristic timescale for structures in the ISM. It is given by:
\begin{equation}
	\label{eq:t_ff}
	t_\mathrm{ff} = \sqrt{\frac{3\pi}{32 G \rho_0}}
\end{equation}
Given that we are interested in the current infall velocity of a set of clumps, we have to assume some initial density for the clump prior to its collapse into its present state. We use the typical average number density for a molecular cloud of $n_0=\qty{3e2}{\per\cc}$ \citep{Peretto2023} as our initial density (corresponding to a $\rho_0 \sim \qty{17}{\Msol\per\cubic\parsec}$, with a mean molecular mass $\mu=2.33$). This yields a free-fall time of around $\qty{1.95}{\mega\yr}$. 

Under the assumption that mass is conserved during the collapse, and thus that no mass is being accreted onto the clump, we can express the initial radius $r_0$ of the clump as:
\begin{equation}
	\label{eq:r_init}
	r_0 = r \left(\frac{n}{n_0}\right)^{1/3}
\end{equation}
where $r$ and $n$ are the current clump radius and mean number density. We could potentially use the free-fall time and the change in clump radius to estimate the infall velocity, but doing so would lead to  two main issues. Firstly, calculating $r_0/t_\mathrm{ff}$ will only give an average collapse velocity, and not be representative of the current infall velocity as the rate of collapse is not constant, but starts slowly and quickly ramps up as you approach $t=t_\mathrm{ff}$ \citep[see Figure 6]{Girichidis2014}. Secondly, the free-fall time is the time it takes for a parcel of gas to collapse to a single point, and not to the current radius $r$ of the clump. 

Instead, \cite{Girichidis2014} define the dimensionless time $\tau_\mathrm{frac}=t/t_\mathrm{ff}$, which is effectively the fraction of the free-fall time that has elapsed since $t=0$. A close approximation of this quantity can be expressed in terms of radius as:
\begin{equation}
	\label{eq:tau_factor}
	\tau_\mathrm{frac} = \sqrt{1 - \left(\frac{r}{r_0}\right)^{3/a}}
\end{equation}
where $a=1.8614$ is a numerical factor that minimises the error between this approximation and the exact form of $\tau_\mathrm{frac}$.  We can obtain an equation for the time evolution of the clump radius by rearranging \eqref{eq:tau_factor} for $r$ and substituting $\tau_\mathrm{frac}=t/t_\mathrm{ff}$:
\begin{equation}
	\label{eq:r_time}
	r(t) = r_0 \left(1 - \frac{t^2}{t_\mathrm{ff}^2}\right)^{a/3}
\end{equation}
We can then take the time derivative of this equation to obtain the time evolution of the infall velocity:
\begin{equation}
	\label{eq:drdt}
	\frac{\mathrm{d}r}{\mathrm{d}t} = \frac{-2 a r_0}{3} \frac{t}{t_\mathrm{ff}^2} \left(1-\frac{t^2}{t_\mathrm{ff}^2}\right)^{(a/3) - 1}
\end{equation}
and substitute $t = \tau_\mathrm{frac} \cdot t_\mathrm{ff}$ so that we can more simply calculate the infall velocity $v_\mathrm{inf}$ as a function of the initial radius, free-fall time and $\tau_\mathrm{frac}$:
\begin{equation}
	\label{eq:v_inf}
	v_\mathrm{inf} = \frac{-2 a r_0}{3} \frac{\tau_\mathrm{frac}}{t_\mathrm{ff}} (1-\tau_\mathrm{frac}^2)^{(a-3)/3}
\end{equation}
Assuming that the clump remains spherically symmetric over the course of its collapse, the average clump density as a function of time can be expressed as:
\begin{equation}
	\label{eq:n_tau}
	n(t) = n_0 \left(1 - \frac{t^2}{t_\mathrm{ff}^2}\right)^{-a}
\end{equation}

\figref{fig:vinf} shows the resulting evolution of $v_\mathrm{inf}$ as a function of density and time for a sample of clumps.
\newpage
\onecolumn

\section{3D visualisations of the fit results} 
\label{app:3d_visualisations_of_the_fit_results}

\begin{figure*}
	\centering
	\vspace{-5mm}
	\subfloat{
	\includegraphics[width=0.48\textwidth]{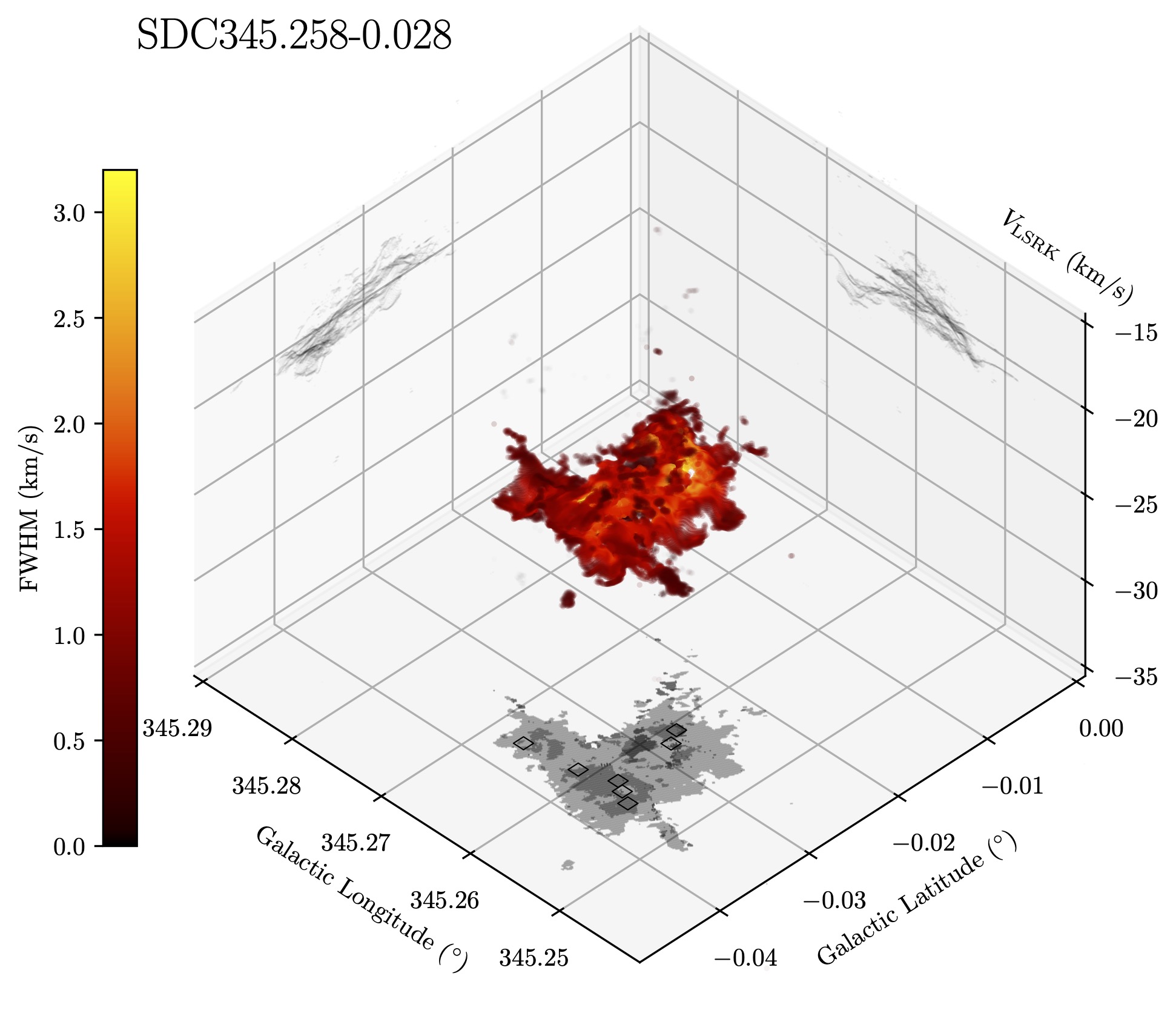}
	}
	\subfloat{
	\includegraphics[width=0.48\textwidth]{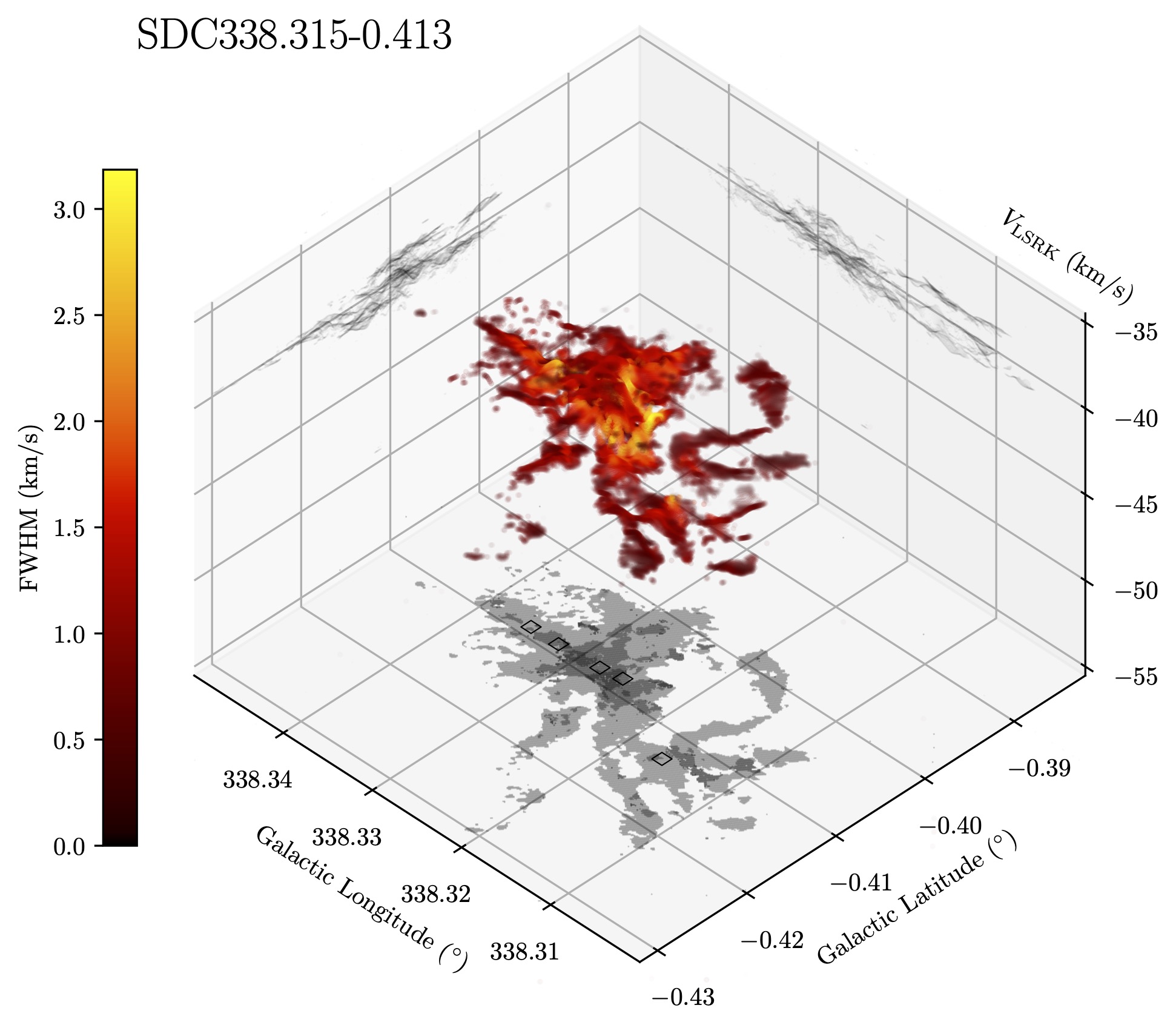}
	}\\[-4mm]
	\subfloat{
	\includegraphics[width=0.48\textwidth]{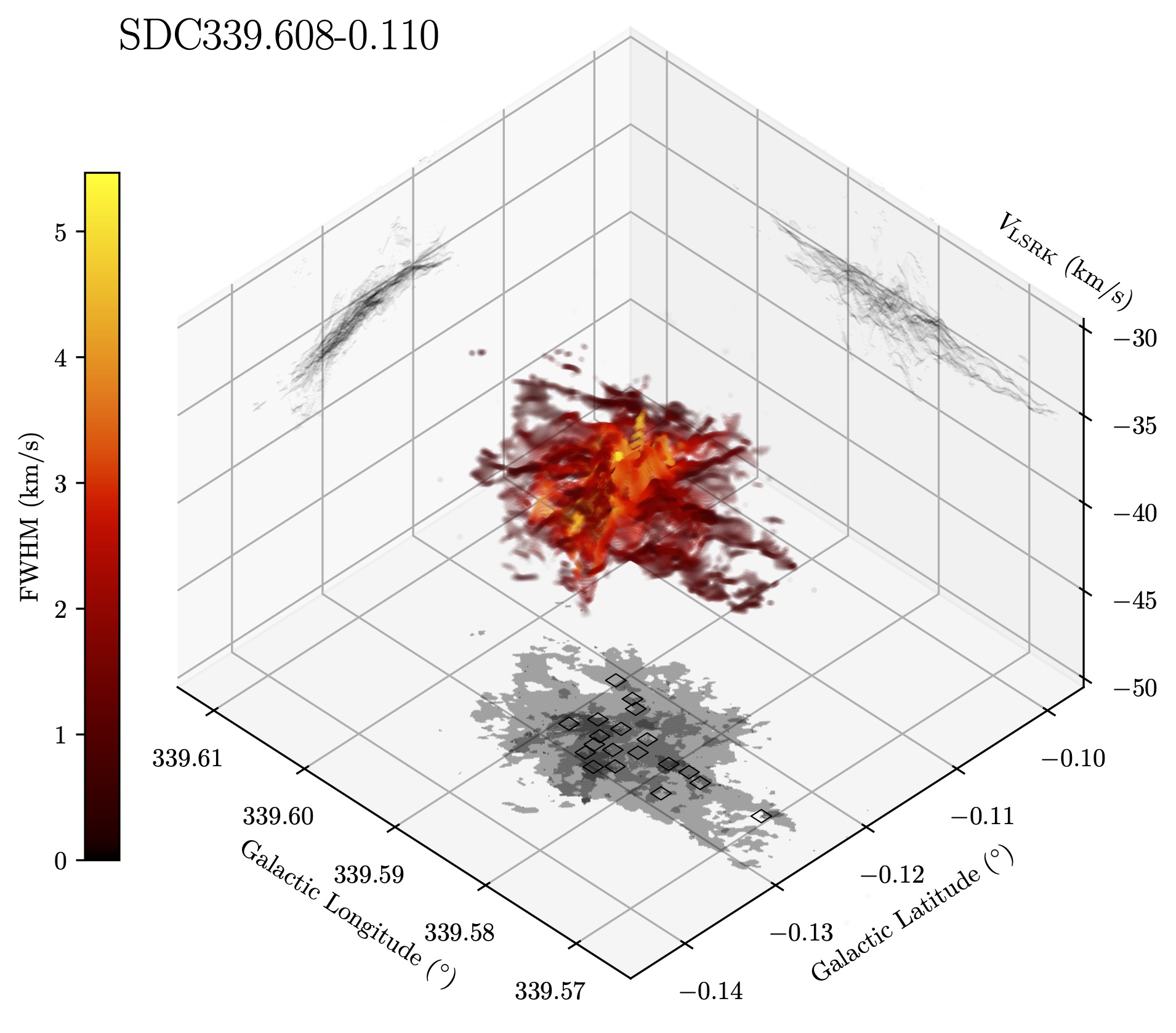}
	}
	\subfloat{
	\includegraphics[width=0.48\textwidth]{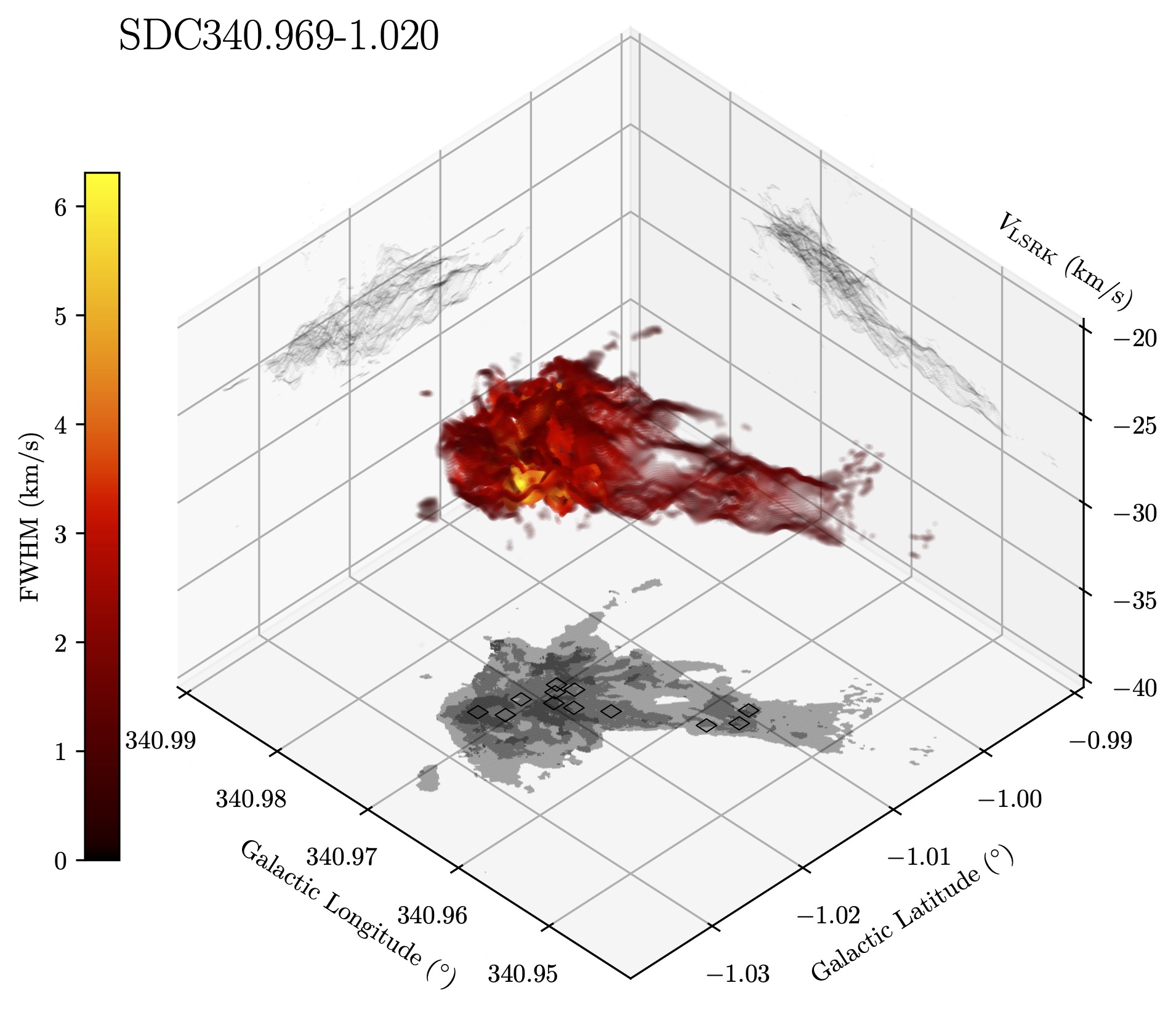}
	}\\[-4mm]
	\subfloat{
	\includegraphics[width=0.48\textwidth]{3D_fwhm_SDC326.jpg}
	}
	\subfloat{
	\includegraphics[width=0.48\textwidth]{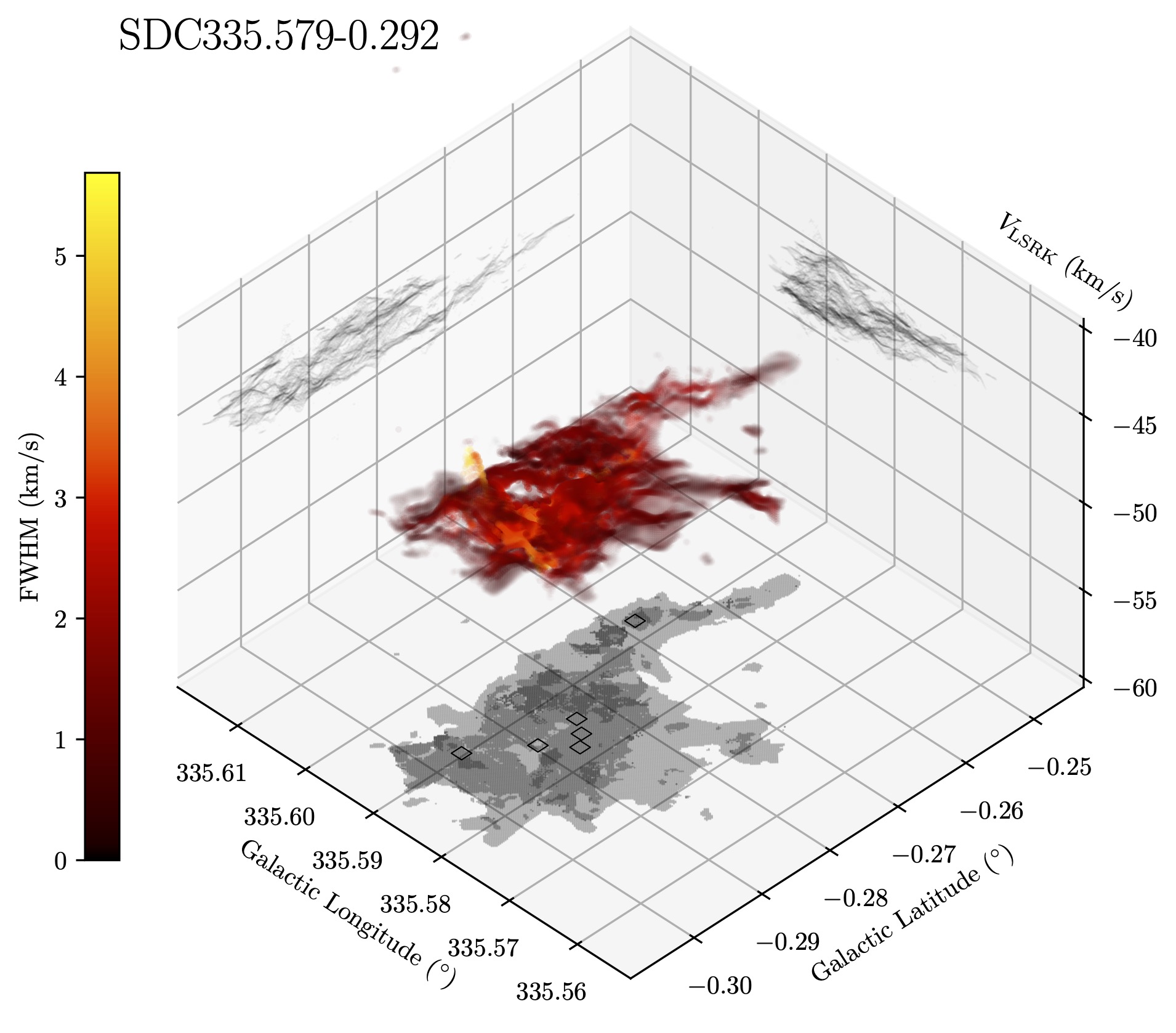}
	}
	\caption{3D PPV ($l$, $b$, $v$) visualisations of the fit results produced by \mwydyn{}. The points are coloured by the best fit FWHM value. On the lower box surface is a 2D projection ($l$, $b$) showing the number of fitted velocity components along the line of sight, with darker grey indicating more components. Core positions are marked with boxes. The left and right surfaces show 2D projections in both ($b$, $v$) and ($l$, $v$), respectively.
	}\label{fig:PPV_3D}
\end{figure*}

\newpage
\onecolumn

\section{Distributions of fit properties with radius}
\label{app:distributions_of_fit_properties_with_radius}

\begin{figure*}
	\centering
	\includegraphics[width=0.85\textwidth]{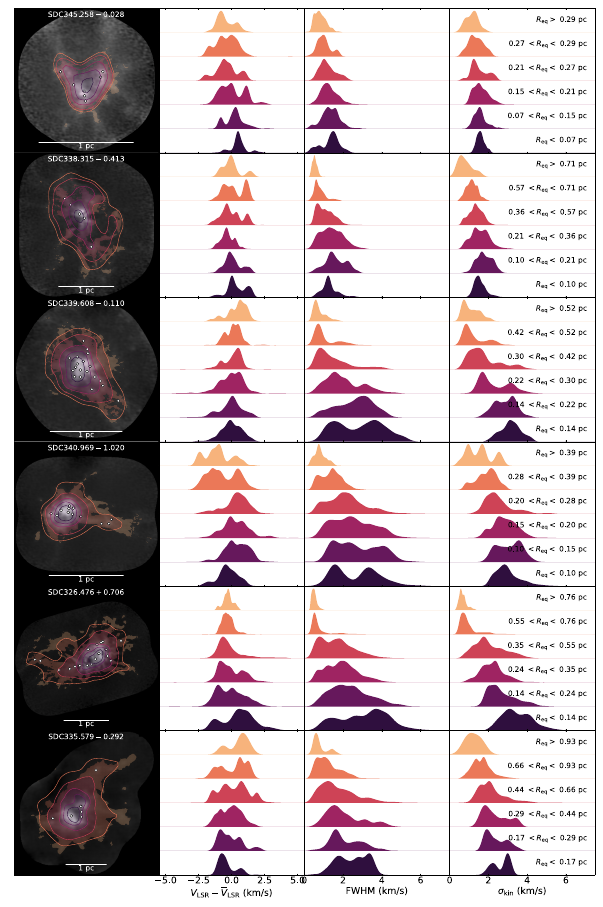}
		\caption{\emph{From left to right}: \nthp{} integrated intensity images for each cloud, with \emph{Herschel} column density contours (same as \figref{fig:ncomp_radius}), and then the KDEs of the fitted velocity centroids, FWHM values, and the total velocity dispersions colour-coded by their respective contours in the image on the left. The points that do not fall within the outermost contour level are assigned to the topmost distribution. All of these distributions are integrated intensity weighted.
		}
	\label{fig:kde_radius}
\end{figure*}

\newpage
\onecolumn

\section{Self absorption in SDC335}
\label{sec:self_absorption_in_sdc335}

\begin{figure*}
	\centering
	\includegraphics[width=0.95\textwidth]{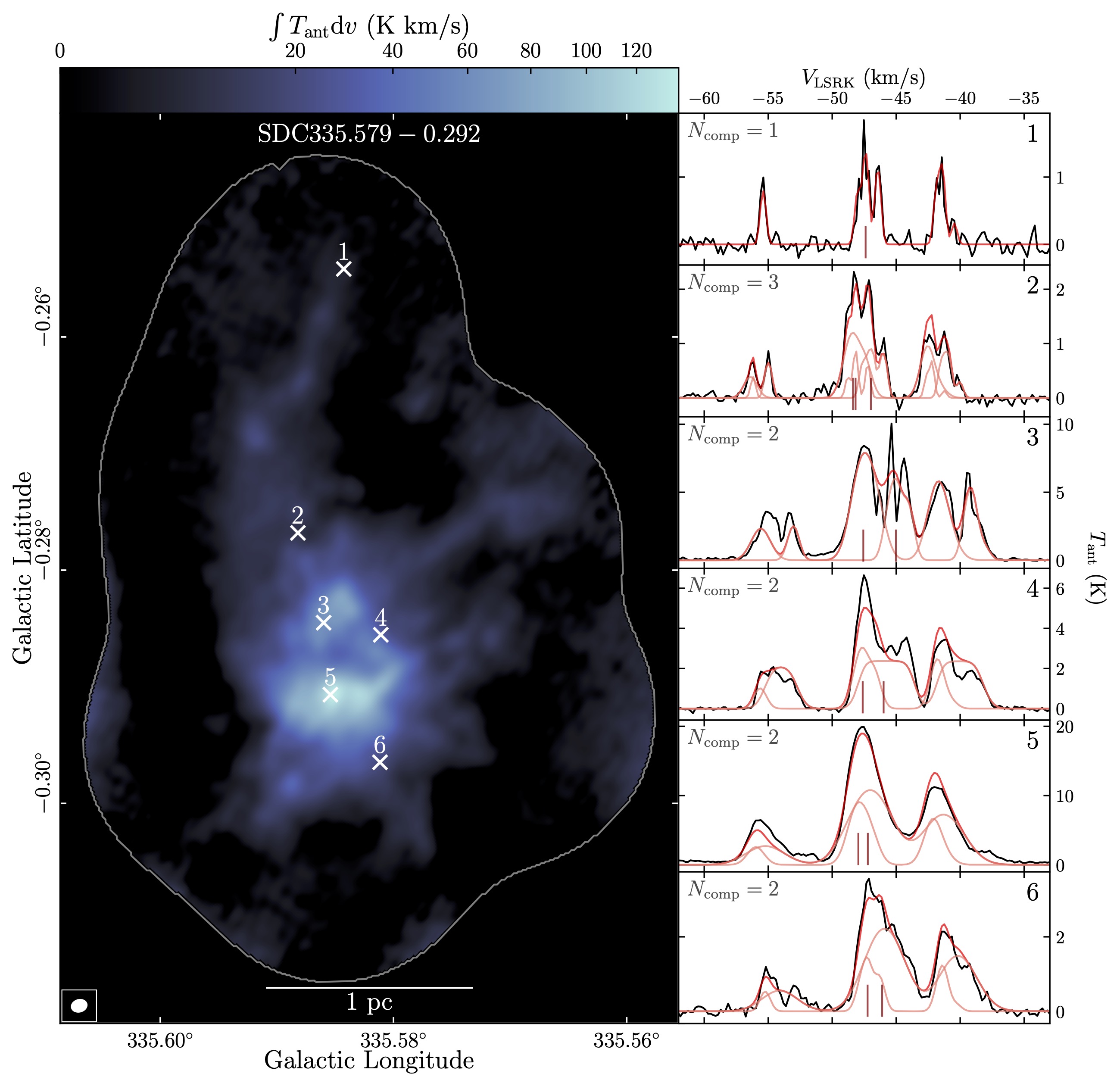}
		\caption{Integrated intensity \nthp(J=1--0) image of SDC335, with example spectra in black (numbered 1--6, with locations marked on the map), and their corresponding best fitting models produced by \mwydyn{} in red.
		}
	\label{fig:SDC335_example}
\end{figure*}

\newpage
\onecolumn

\section{Position-velocity diagrams}
\label{app:position_velocity_diagrams}

\begin{figure*}
	\centering
	\subfloat{
	\includegraphics[width=0.99\textwidth]{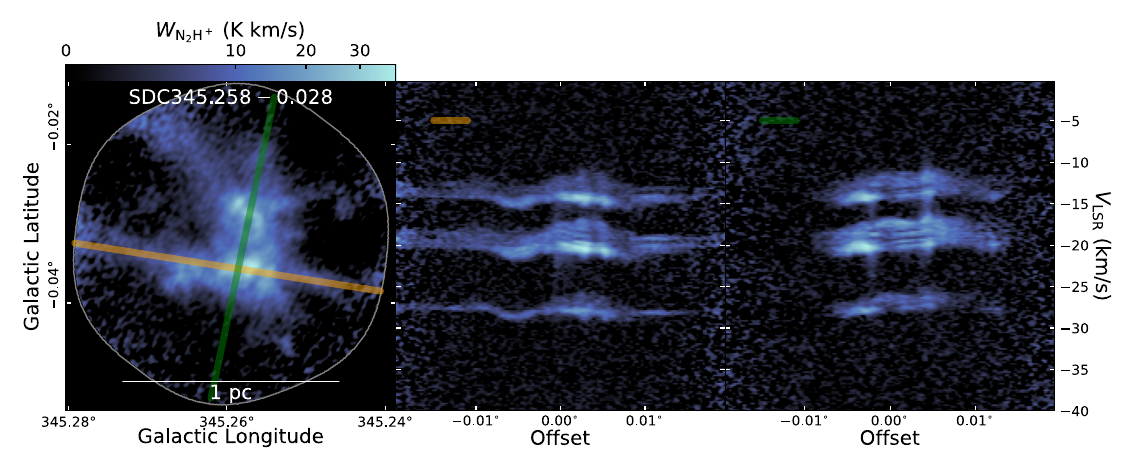}
	}\\
	\vspace{-0cm}
	\subfloat{
	\includegraphics[width=0.99\textwidth]{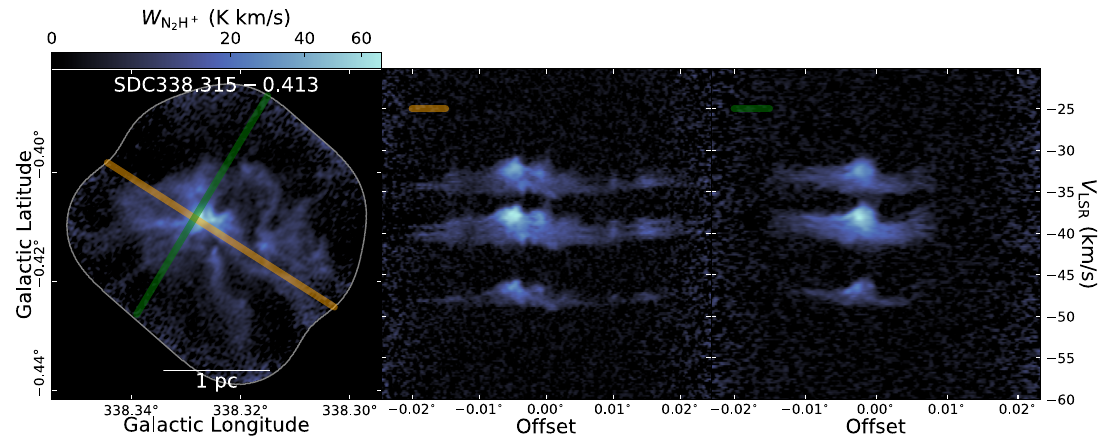}
	}
	\caption{The left-hand panels show the integrated \nthp(J=1--0) intensity data for each hub,  with the orange and green line indicating the axes for which the PV-diagram were generated, which are shown in the middle (orange) and right-hand (green) panels.
		}
	\label{fig:pv_diagrams}
\end{figure*}

\begin{figure*}\ContinuedFloat
	\centering
	\subfloat{
	\includegraphics[width=0.99\textwidth]{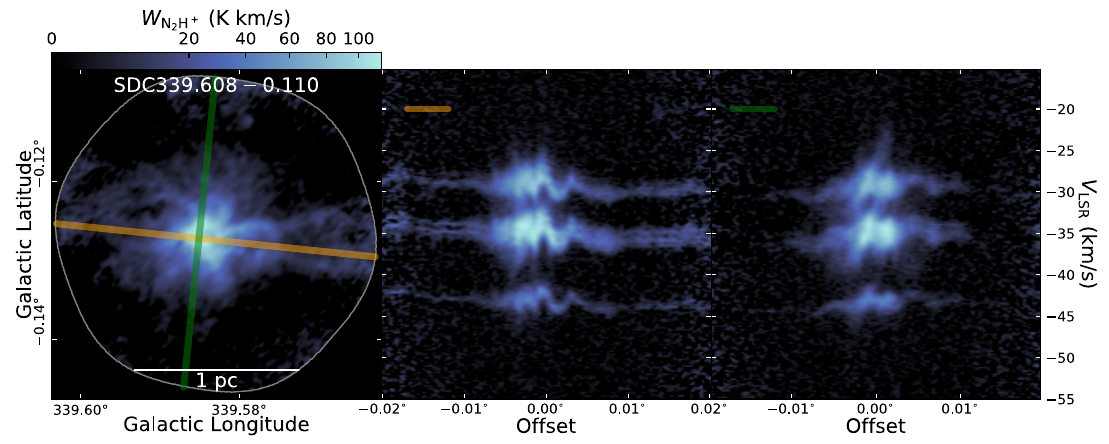}
	}\\
	\subfloat{
	\includegraphics[width=0.99\textwidth]{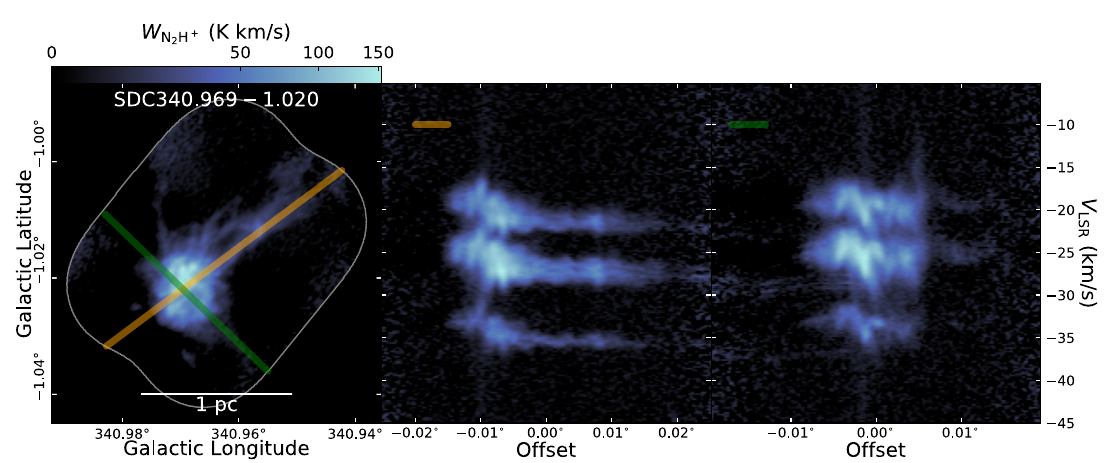}
	}
	\caption[]{\emph{(continued)}}
\end{figure*}

\begin{figure*}\ContinuedFloat
	\centering
	\subfloat{
	\includegraphics[width=0.99\textwidth]{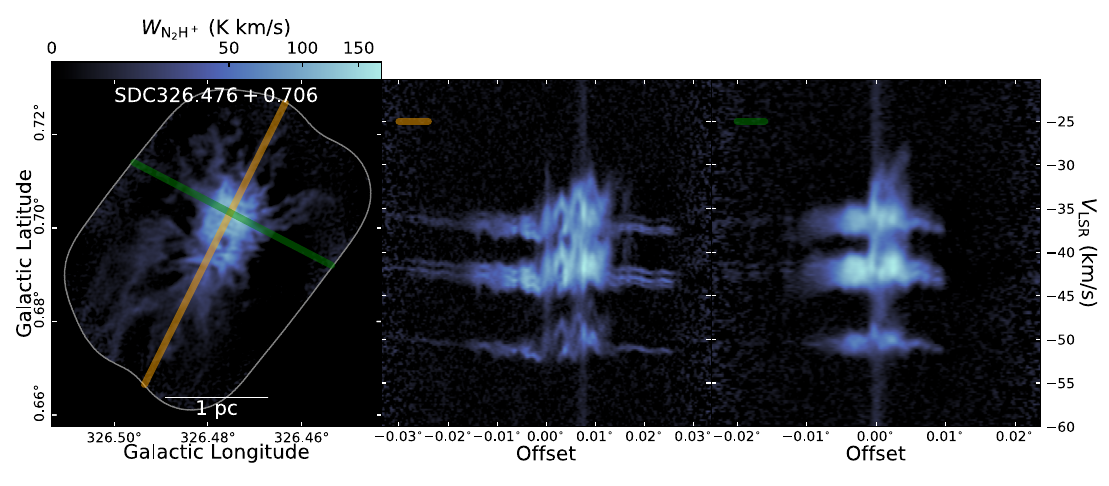}
	}\\
	\subfloat{
	\includegraphics[width=0.99\textwidth]{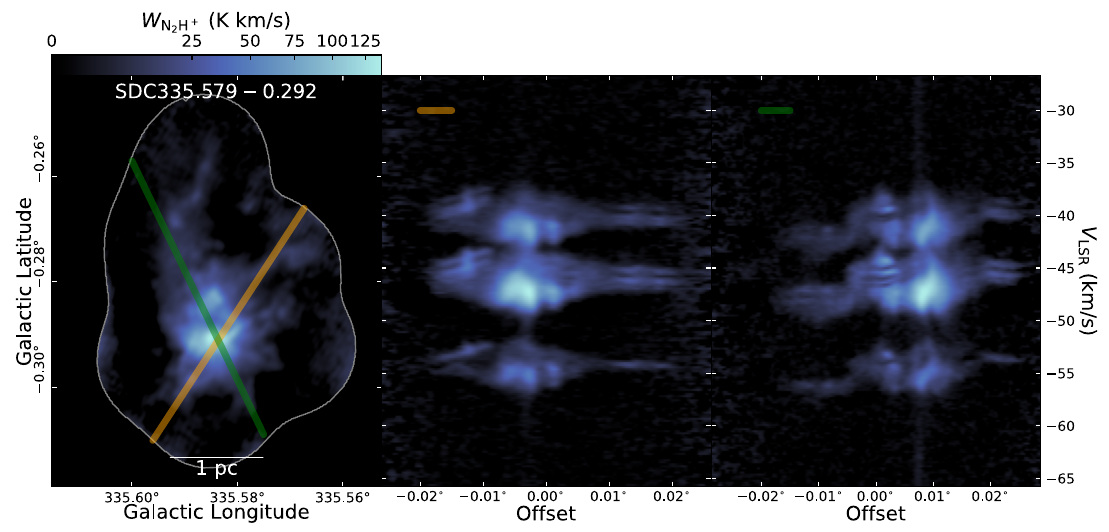}
	}
	\caption[]{\emph{(continued)}}
\end{figure*}


\bsp	
\label{lastpage}
\end{document}

%% file: obs_stats.tex
\begin{table}
	\caption{The 6 HFSs we observed with ALMA and a summary of observational properties of of our combined TP+7\si{~\m}+12\si{~\m} observations of \nthp(J=1--0). The HFSs will hereafter be referred to by their shorthand names highlighted in bold. The RMS noise varies between spectra in each field, so here it represents the mean RMS for the whole cube. The linear resolution $x_d$ corresponds to the physical size of the synthesised beam major axis at the source distance.
	}
	\label{tab:obs_prop}
	\centering
	\begin{tabular}{c c c c c}
		\hline\hline
		SDC Name & Beam & PA & RMS & $x_d$ \\
		\citep{Peretto2009} & ($\si{\arcsecond} \times \si{\arcsecond}$) & (\si{\degree}) & (\si[per-mode=symbol]{\kelvin}) & (\si{\parsec}) \\
		\hline
		\textbf{SDC326}.476+0.706  & $3.26 \times 2.49$ & $ 67.42$ & $0.15$ & 0.041 \\
		\textbf{SDC335}.579--0.292 & $5.21 \times 4.16$ & $-74.50$ & $0.08$ & 0.082 \\
   		\textbf{SDC338}.315--0.413 & $3.38 \times 2.25$ & $ 80.47$ & $0.16$ & 0.048 \\
   		\textbf{SDC339}.608--0.113 & $3.36 \times 2.21$ & $ 81.95$ & $0.18$ & 0.045 \\
		\textbf{SDC340}.969--1.020 & $3.36 \times 2.23$ & $ 80.73$ & $0.17$ & 0.040 \\
		\textbf{SDC345}.258--0.028 & $3.29 \times 2.17$ & $ 81.43$ & $0.18$ & 0.033 \\
		\hline
	\end{tabular}
\end{table}

%% file: sample_stats.tex
\begin{table}
	\caption[Summary of updated source properties.]{List of sources and their properties from \protect\cite{Anderson2021} (with superscript A+21), and updated properties in this work. The clumps have been ordered by ascending clump mass.}
	\label{tab:sample}
	\centering
	\begin{tabular}{c c c c c c}
		\hline\hline
		SDC Name & $d$ & $M_{\mathrm{clump}}^{A+21}$ & $f_\mathrm{MMC}^{A+21}$ & $M_{\mathrm{clump}}$ & $f_\mathrm{MMC}$ \\
		\citep{Peretto2009} & (\si{\parsec}) & (\si{\Msol}) & (\%) & (\si{\Msol}) & (\%) \\
		\hline
		SDC345.258--0.028 & 2090 & 135 & 11.0 & 285 & 8.1 \\
		SDC338.315--0.413 & 2940 & 213 & 12.4 & 672 & 3.9 \\
		SDC339.608--0.113 & 2740 & 942 & 3.2 & 1150 & 2.6 \\
		SDC340.969--1.020 & 2210 & 1768 & 7.0 & 1622 & 7.6 \\
		SDC326.476+0.706  & 2610 & 2399 & 22.2 & 3307 & 16.1 \\
		SDC335.579--0.292 & 3230 & 3739 & 24.4 & 4926 & 18.5 \\
		\hline
	\end{tabular}
\end{table}

%% file: n2h+_model.tex
\begin{table}
	\caption{Parameters of the \nthp(J=1--0) multiplet used in our fitting model. The velocities are with respect to the reference component, shown in bold, for which our fitted centroid velocities correspond to.}
	\label{tab:multiplet}
	\centering
	\begin{tabular}{c c c c}
		\hline\hline
		Transition & $\nu_i$ & $r_i$ & $\delta v_i$ \\
		($J F_1 F \rightarrow J' {F_1}' F'$) & (\si{\mega\hertz}) &  & (\si{\kilo\meter\per\second}) \\
		\hline
		$1\ 1\ 0 \rightarrow 0\ 1\ 1$ & $93171.6086$          & $1/27$ & $6.9360$  \\
		$1\ 1\ 2 \rightarrow 0\ 1\ 2$ & $93171.9054$          & $5/27$ & $5.9841$  \\
		$1\ 1\ 1 \rightarrow 0\ 1\ 0$ & $93172.0403$          & $3/27$ & $5.5452$  \\
		$1\ 2\ 2 \rightarrow 0\ 1\ 1$ & $93173.4675$          & $5/27$ & $0.9560$  \\
		$\mathbf{1\ 2\ 3 \rightarrow 0\ 1\ 2}$ & $\mathbf{93173.7643}$ & $\mathbf{7/27}$ & $\mathbf{0.0000}$  \\
		$1\ 2\ 1 \rightarrow 0\ 1\ 1$ & $93173.9546$          & $3/27$ & $-0.6109$ \\
		$1\ 0\ 1 \rightarrow 0\ 1\ 2$ & $93176.2527$          & $3/27$ & $-8.0064$ \\
		\hline
	\end{tabular}
\end{table}